\documentclass[aps,prl,twocolumn,amsmath,amssymb,tightenlines,epsfig,floatfix,superscriptaddress]{revtex4-1}
\usepackage{graphicx}
\usepackage{dcolumn}	
\usepackage{bm}			
\usepackage{amsfonts}
\usepackage{xspace}
\usepackage{color}
\usepackage{epstopdf}
\usepackage{multirow}

\begin{document}

\newcommand{\psihat}{\ensuremath{\hat{\psi}}\xspace}
\newcommand{\psihatd}{\ensuremath{\hat{\psi}^{\dagger}}\xspace}
\newcommand{\ahat}{\ensuremath{\hat{a}}\xspace}
\newcommand{\Ham}{\ensuremath{\mathcal{H}}\xspace}
\newcommand{\ahatd}{\ensuremath{\hat{a}^{\dagger}}\xspace}
\newcommand{\bhat}{\ensuremath{\hat{b}}\xspace}
\newcommand{\bhatd}{\ensuremath{\hat{b}^{\dagger}}\xspace}
\newcommand{\boldr}{\ensuremath{\mathbf{r}}\xspace}
\newcommand{\dr}{\ensuremath{\,d^3\mathbf{r}}\xspace}
\newcommand{\dk}{\ensuremath{\,d^3\mathbf{k}}\xspace}
\newcommand{\etal}{\emph{et al.\/}\xspace}
\newcommand{\ie}{i.e.}
\newcommand{\eq}[1]{Eq.~(\ref{#1})\xspace}
\newcommand{\fig}[1]{Fig.~\ref{#1}\xspace}
\newcommand{\abs}[1]{\left| #1 \right|}
\newcommand{\proj}[2]{\left| #1 \rangle\langle #2\right| \xspace}
\newcommand{\Qhat}{\ensuremath{\hat{Q}}\xspace}
\newcommand{\Qhatd}{\ensuremath{\hat{Q}^\dag}\xspace}
\newcommand{\phihatd}{\ensuremath{\hat{\phi}^{\dagger}}\xspace}
\newcommand{\phihat}{\ensuremath{\hat{\phi}}\xspace}
\newcommand{\boldk}{\ensuremath{\mathbf{k}}\xspace}
\newcommand{\boldp}{\ensuremath{\mathbf{p}}\xspace}
\newcommand{\boldsigma}{\ensuremath{\boldsymbol\sigma}\xspace}
\newcommand{\boldalpha}{\ensuremath{\boldsymbol\alpha}\xspace}
\newcommand{\grad}{\ensuremath{\boldsymbol\nabla}\xspace}
\newcommand{\parti}[2]{\frac{ \partial #1}{\partial #2} \xspace}
 \newcommand{\vs}[1]{\ensuremath{\boldsymbol{#1}}\xspace}
\renewcommand{\v}[1]{\ensuremath{\mathbf{#1}}\xspace}
\newcommand{\Psihat}{\ensuremath{\hat{\Psi}}\xspace}
\newcommand{\Psihatd}{\ensuremath{\hat{\Psi}^{\dagger}}\xspace}
\newcommand{\Vhatd}{\ensuremath{\hat{V}^{\dagger}}\xspace}
\newcommand{\Xhat}{\ensuremath{\hat{X}}\xspace}
\newcommand{\Xhatd}{\ensuremath{\hat{X}^{\dag}}\xspace}
\newcommand{\Yhat}{\ensuremath{\hat{Y}}\xspace}
\newcommand{\Yhatd}{\ensuremath{\hat{Y}^{\dag}}
\xspace}
\newcommand{\jhat}{\ensuremath{\hat{J}}
\xspace}
\newcommand{\lhat}{\ensuremath{\hat{L}}
\xspace}
\newcommand{\Nhat}{\ensuremath{\hat{N}}
\xspace}
\newcommand{\ddt}{\ensuremath{\frac{d}{dt}}
\xspace}
\newcommand{\nset}{\ensuremath{n_1, n_2,\dots, n_k}
\xspace}
\newcommand{\sah}[1]{{\color{red}#1}}
\newcommand{\rob}[1]{{\color{blue}#1}}
\newcommand{\sss}[1]{{\color{blue}#1}}

\title{Pumped-Up SU(1,1) Interferometry}
\author{Stuart S.~Szigeti}
\affiliation{School of Mathematics and Physics,  University of Queensland, Brisbane, Queensland 4072, Australia}
\affiliation{ARC Centre of Excellence for Engineered Quantum Systems, University of Queensland, Brisbane, Queensland 4072, Australia}
\author{Robert J.~Lewis-Swan}
\affiliation{School of Mathematics and Physics,  University of Queensland, Brisbane, Queensland 4072, Australia}
\author{Simon A.~Haine}
\affiliation{Department of Physics and Astronomy, University of Sussex, Brighton BN1 9QH, United Kingdom}

\begin{abstract}

Although SU(1,1) interferometry achieves Heisenberg-limited sensitivities, it suffers from one major drawback: Only those particles outcoupled from the pump mode contribute to the phase measurement. Since the number of particles outcoupled to these ``side modes'' is typically small, this limits the interferometer's \emph{absolute} sensitivity. We propose an alternative ``pumped-up'' approach where all the input particles participate in the phase measurement, and show how this can be implemented in spinor Bose-Einstein condensates and hybrid atom-light systems - both of which have experimentally realized SU(1,1) interferometry. We demonstrate that pumped-up schemes are capable of surpassing the shot-noise limit with respect to the total number of input particles and are never worse than conventional SU(1,1) interferometry. Finally, we show that pumped-up schemes continue to excel - both absolutely and in comparison to conventional SU(1,1) interferometry - in the presence of particle losses, poor particle-resolution detection, and noise on the relative phase difference between the two side modes. Pumped-up SU(1,1) interferometry therefore pushes the advantages of conventional SU(1,1) interferometry into the regime of high absolute sensitivity, which is a necessary condition for useful quantum-enhanced devices.

\end{abstract}

\maketitle

\begin{figure}
\includegraphics[width=\columnwidth]{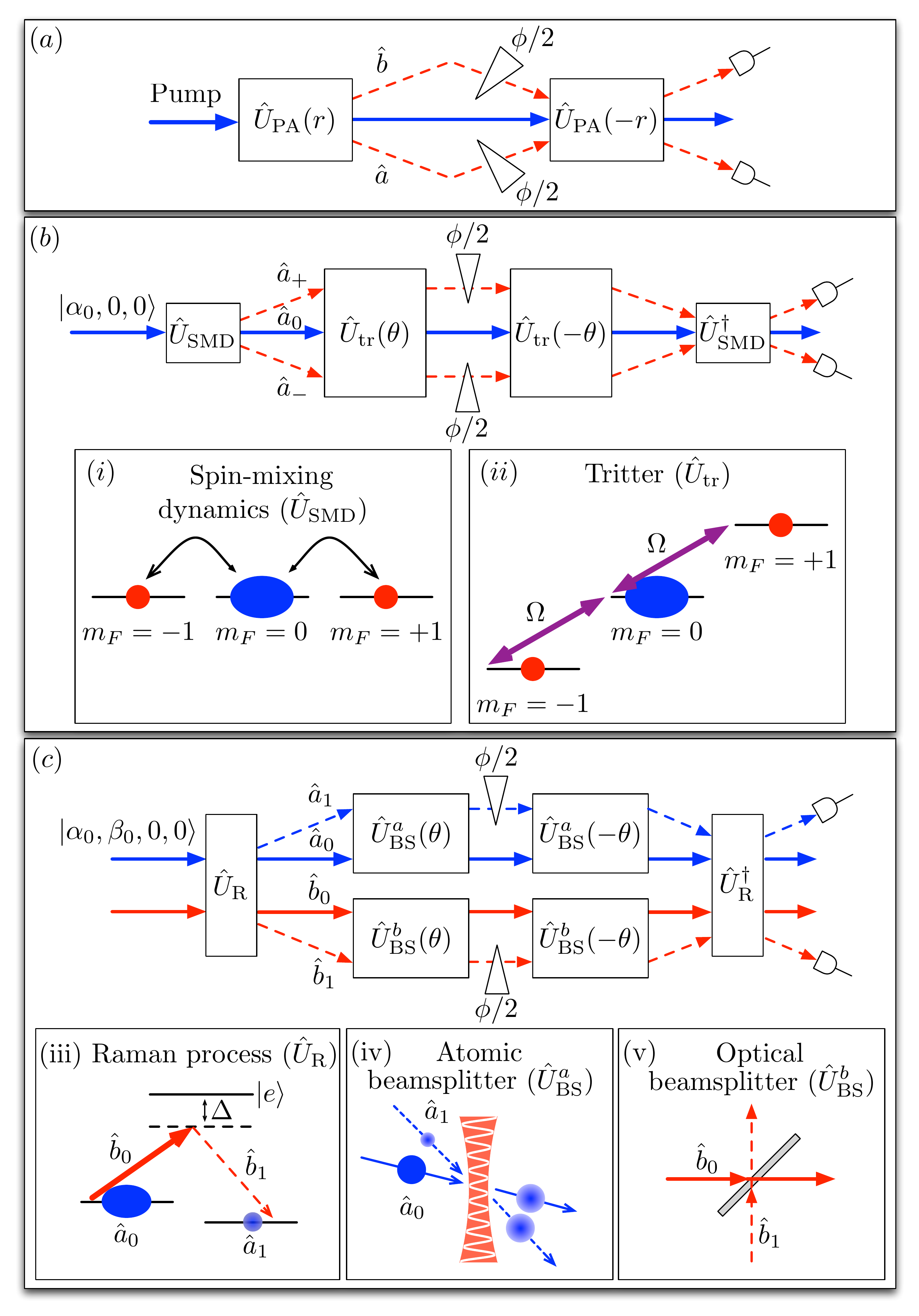}
\caption{(a) A conventional SU(1,1) interferometer, constructed with two active nonlinear beamspliters $\hat{U}_\text{PA}(r)$. (b) Pumped-up SU(1,1) interferometry with three modes of a spinor BEC. Initially, all atoms are in the $m_F = 0$ pump mode, assumed to be a coherent state $|\alpha_0\rangle$ with $|\alpha_0|^2 = \overline{N}$. The active beam splitter $\hat{U}_\text{SMD}$ is achieved via spin-mixing collisions between three hyperfine levels [see (i) and Eq.~(\ref{Ham_SMD})], whereas the pump is mixed with the two side modes using a tritter $\hat{U}_\text{tr}(\theta)$, engineered with coherent radio frequency pulses [see (ii) and Eqs.~(\ref{HP_tritter})]. (c) Pumped-up SU(1,1) interferometry with the four modes of a hybrid atom-light system. The initial pump modes are coherent states, the active beam splitter $\hat{U}_\text{R}(r)$ is realized by FWM engineered with a Raman process [see (iii) and Eq.~(\ref{Ham_FWM})], and pump enhancement is achieved with atomic and optical beam splitters that separately mix the atomic and photonic modes [see (iv) and (v), respectively].
}
\label{fig_scheme}
\end{figure}

Quantum correlations allow precision interferometric measurements below the shot-noise limit \cite{Caves:1981, Toth:2014}. This can be achieved by replacing the input state of a conventional interferometer with a nonclassical state; this is the approach being pursued in gravitational wave detection \cite{LIGO:2011, LIGO:2013}, where the vacuum port of a Michelson interferometer is substituted for a squeezed-light source. 
Unfortunately, the fragility of highly correlated quantum states to detection losses severely limits the quantum enhancement achievable in practice \cite{Demkowicz-Dobrzanski:2012}. 
An alternative approach is to design an interferometer where the quantum correlations are generated within the interferometer, thereby making it robust to these losses. The archetypical example is a SU(1,1) interferometer \cite{Yurke:1986, Leonhardt:1994}, which is configured as a Mach-Zehnder with the passive beam splitters replaced by \emph{active} nonlinear beam splitters that create or annihilate pairs of correlated particles [see Fig.~\ref{fig_scheme}(a)]. This generates a high degree of particle entanglement \emph{within} the interferometer, allowing phase measurements at the ultimate Heisenberg limit while additionally providing a robustness to inefficient particle detection \cite{Marino:2012, Ou:2012}. This excellent ``per particle'' sensitivity and robustness has resulted in a strong theoretical interest in SU(1,1) interferometry \cite{Gabbrielli:2015, Chen:2016, Gong:2016, Xin:2016}, and its experimental realization in optical systems \cite{Jing:2011,Hudelist:2014}, hybrid atom-light interferometers \cite{Chen:2015b}, and spinor Bose-Einstein condensates (BECs) \cite{Gross:2010, Peise:2015, Linnemann:2016}.

Unfortunately, the prospect of a \emph{high-precision} SU(1,1) interferometer is limited. In practice, it is difficult to engineer nonlinear active beam splitters that are both reversible and capable of outcoupling even modest numbers of particles. 
For example, the Heisenberg-limited phase measurement reported in \cite{Linnemann:2016} was made with a mere $2.8 \pm 0.2$ particles on average. Consequently, the promise of Heisenberg-limited sensitivities is of little practical benefit, especially when sophisticated \emph{classical} interferometers display superior absolute sensitivities by many orders of magnitude and suffer none of the robustness issues that afflict quantum-enhanced devices. The crux of the issue is that SU(1,1) interferometry is inherently wasteful; it requires the generation and manipulation of large numbers of particles but does not make use of all these particles within the phase measurement. As a general heuristic, a necessary condition for a high-precision (i.e., \emph{useful}) quantum-enhanced device is that the quantum enhancement provide additional sensitivity beyond the shot-noise limit with respect to the \emph{total} particle number. 

In this Letter we present a modification to SU(1,1) interferometry that (a) uses \emph{all} particles to make the phase measurement, (b) gives sub-shot-noise sensitivities with respect to the total particle number, and (c) is surprisingly more robust than conventional SU(1,1) interferometry to inefficient particle detection. Our ``pumped-up'' approach linearly mixes the correlated pairs of particles with the pump mode(s) from which these particles are outcoupled and, therefore, represents only a small increase in the complexity of the interferometer design. Nevertheless, pumped-up SU(1,1) interferometry is, in principle, never worse than conventional SU(1,1) interferometry, and is usually orders of magnitude more sensitive, even in the presence of typical losses. We illustrate the general principles of pumped-up SU(1,1) interferometry by considering specific implementations in (i) spinor BECs and (ii) hybrid atom-light systems. Both platforms have experimentally realized proof-of-principle SU(1,1) interferometry \cite{Chen:2015b, Linnemann:2016} and, therefore, represent strong candidate systems for implementing our pumped-up approach. 

\emph{Conventional SU(1,1) interferometry.---}The first beam splitter in a SU(1,1) interferometer actively creates correlated particle pairs via parametric amplification, described by the unitary $\hat{U}_\text{PA}(r) = \exp[-i r (\hat{a}_1^\dag \hat{a}_2^\dag + \hat{a}_1 \hat{a}_2)]$, where $\hat{a}_1$ and $\hat{a}_2$ are the two bosonic modes that form the arms of the interferometer (the ``side modes''). Since these modes are initially vacuum, this unitary produces a two-mode squeezed vacuum state - which is a coherent superposition of twin-Fock states - with average particle number $\mathcal{N}_s \equiv 2 \sinh^2 r$ \cite{Loudon:1987}. These particles are assumed to be outcoupled from an undepleted reservoir (the ``pump mode''), whose average occupation is much larger than $\mathcal{N}_s$. After some interrogation time, which imprints a phase $\phi/2$ on each side mode, a second parametric amplifier reverses the first [see Fig.~1(a)]; this is conveniently achieved by imposing a $\pi/2$ phase shift on the pump such that $r \to -r$. A measurement of the number sum of the two side modes $\hat{N}_s = \hat{a}_1^\dag \hat{a}_1 + \hat{a}_2^\dag \hat{a}_2$ at the output is sensitive to the phase $\phi$. Explicitly, at the optimal operating point $\phi = 0$, the phase sensitivity of this measurement is Heisenberg limited with respect to $\mathcal{N}_s$:
\begin{equation}
	\Delta \phi_\text{SU(1,1)} = \frac{\sqrt{\text{Var}(\hat{N}_s)}}{|\partial \langle \hat{N}_s \rangle / \partial \phi|} \Bigg|_{\phi = 0} = \frac{1}{\sqrt{\mathcal{N}_s(\mathcal{N}_s+2)}}.
\end{equation} 

We consider two physical systems which have experimentally realized SU(1,1) interferometry:
	
\emph{(i) Spinor BEC}: The hyperfine manifold of a spin-1 BEC of ultracold atoms can be used to construct an effective three-level system. Spin-mixing collisions coherently outcouple pairs of atoms from the $m_F = 0$ state (pump mode $\hat{a}_0$) to the $m_F = \pm 1$ states (side modes $\hat{a}_\pm$) [see Fig.~\ref{fig_scheme}(i)]. The full spin-mixing dynamics are given by \cite{Law:1998, Lewis-Swan:2013}
\begin{align}
	\hat{H}_\text{SMD} 	&=  \hbar \kappa [ \hat{a}_0^2 \hat{a}_+^\dag \hat{a}_-^\dag + (\hat{a}_0^\dag)^2 \hat{a}_+ \hat{a}_- ]  \notag \\
					&+ \hbar \kappa ( \hat{N}_0 - \tfrac{1}{2} ) ( \hat{N}_+ + \hat{N}_- ) + \hbar q ( \hat{N}_+ + \hat{N}_- ), \label{Ham_SMD}
\end{align} 
where $\hat{N}_i \equiv \hat{a}_i^\dag \hat{a}_i$. 
By dynamically tuning $q$ with a magnetic field, the quadratic Zeeman shift (third term) cancels collisional shifts due to s-wave scattering of the three modes (second term) \cite{Zeeman_shift, Linnemann:2016}.
Then, provided $\langle \hat{N}_0 \rangle \gg \langle \hat{N}_\pm \rangle$ throughout the interaction time $t$, the undepleted pump approximation $\hat{a}_0 \to \sqrt{\overline{N}}$ holds (for average total particle number $\overline{N}$), and we realize $\hat{U}_\text{PA}(r)$ with $r = \overline{N} \kappa t$. 

\emph{(ii) Hybrid atom-light system}: Four-wave mixing (FWM) via a Raman pulse generates atom-light entanglement. For an atomic ensemble prepared in pump mode $\hat{a}_0$, a coherent optical pump beam $\hat{b}_0$ transfers atoms from the pump to another atomic mode $\hat{a}_1$, accompanied by the emission of a photon $\hat{b}_1$ [see Fig.~\ref{fig_scheme}(iii)]. Since outcoupling one atom correlates with the production of one photon this realizes correlated atom-light pairs according to \cite{Moore:2000, Hammerer:2010, Haine:2013, Haine:2016a}
\begin{equation}
	\hat{H}_\text{FWM}	= \hbar \kappa ( \hat{a}_0^\dag \hat{b}_0^\dag \hat{a}_1 \hat{b}_1 + \hat{a}_0 \hat{b}_0 \hat{a}_1^\dag \hat{b}_1^\dag). \label{Ham_FWM}
\end{equation}
If both pump modes $\hat{a}_0$ and $\hat{b}_0$ remain highly occupied compared with the side modes $\hat{a}_1$ and $\hat{b}_1$, then the undepleted pump approximation holds ($\hat{a}_0 \to \sqrt{N_{a_0}}$ and $\hat{b}_0 \to \sqrt{N_{b_0}}$ if both pumps are in phase) and we realize $\hat{U}_\text{PA}(r)$ with $r = \sqrt{N_{a_0} N_{b_0}} \kappa t$.

\emph{Pumped-up SU(1,1) interferometry with spinor BECs. ---} We aim to boost the absolute sensitivity of the interferometer by linearly mixing the pump mode $\hat{a}_0$ with side modes $\hat{a}_\pm$ after the first nonlinear beam splitter described by Eq.~(\ref{Ham_SMD}). We do this via a variable-angle three-mode beam splitter (i.e. \emph{tritter}):
\begin{equation}
	\hat{H}_\text{tr} = \tfrac{\hbar \Omega}{\sqrt{2}}\big[ e^{i \vartheta}\hat{a}_0^\dag(\hat{a}_+ + \hat{a}_-) + e^{-i \vartheta} \hat{a}_0(\hat{a}_+^\dag + \hat{a}_-^\dag) \big], \label{tritter_Ham}
\end{equation}
which evolves the modes according to
\begin{subequations}
\label{HP_tritter}
\begin{align}
	\hat{a}_\pm(\theta) 	&= \hat{a}_\pm \cos^2(\tfrac{\theta}{2}) - \hat{a}_\mp \sin^2(\tfrac{\theta}{2}) - \tfrac{i e^{-i \vartheta}}{\sqrt{2}} \hat{a}_0 \sin \theta, \\
	\hat{a}_0(\theta)	&= \hat{a}_0 \cos \theta - \tfrac{i e^{i \vartheta}}{\sqrt{2}} ( \hat{a}_+ + \hat{a}_- ) \sin \theta,
\end{align}
\end{subequations}
where $\theta = \Omega t$ and $\vartheta$ are the tritter angle and phase, respectively. A tritter is achieved by coherently coupling the $m_F = 0$ state to the $m_F = \pm1$ states via a radio frequency pulse of Rabi frequency $\Omega$ and phase $\vartheta$, as illustrated in Fig.~\ref{fig_scheme}(ii). This can be done with high fidelity and on time scales much faster than the nonlinear outcoupling process or phase evolution, as demonstrated experimentally in \cite{Kruse:2016}. After the first tritter, we assume a period of phase evolution that writes a phase $\phi/2$ onto each side mode; the interferometer is then closed by implementing a second tritter (with $\theta \to -\theta$, achievable by changing $\vartheta \to \vartheta + \pi$) and second period of spin mixing [see Fig.~\ref{fig_scheme}(b)]. 

We first quantify the effect of pump enhancement via the quantum Fisher information (QFI), which places a lower bound on the achievable sensitivity $\Delta \phi \geq 1 / \sqrt{\mathcal{F}}$ called the quantum Cram\'er-Rao bound (QCRB) \cite{Braunstein:1994, Toth:2014, Demkowicz-Dobrzanski:2015}. This bound holds irrespective of the specific measurement signal at the output and phase-estimation procedure; here it is entirely determined by the input state, the dynamics of the first spin-mixing operation, and the first tritter (via the angle $\theta$ and phase $\vartheta$). Specifically, within the undepleted pump regime the QFI is \cite{supplemental, Macri:2016, Steel:1998, Blakie:2008, Polkovnikov:2010, Opanchuk:2013, Walls:2008, Gardiner:2004b,Sinatra:2002,Opanchuk:2012,Olsen:2009}
\begin{align}
	\mathcal{F}(\theta)	&= \overline{N} \sin^2 \theta + \tfrac{1}{4}(\overline{N}-\mathcal{N}_s) \mathcal{G}(\mathcal{N}_s, \vartheta) \sin^2(2\theta) \notag \\
					& + \tfrac{1}{2}\mathcal{N}_s \big\{ \mathcal{N}_s + \big[ 3 + (\mathcal{N}_s + 1) \cos^2 \theta \big] \cos^2 \theta\big\}, \label{QFI_tritter}
\end{align}
where $\mathcal{G}(\mathcal{N}_s, \vartheta) \equiv \mathcal{N}_s - \sqrt{\mathcal{N}_s(\mathcal{N}_s+2)} \sin (2 \vartheta)$. For $\theta = 0$ we recover conventional SU(1,1) interferometry with QFI $\mathcal{F}(0) = \mathcal{N}_s ( \mathcal{N}_s + 2)$. Indeed, it trivially follows that $\max_\theta \mathcal{F}(\theta) \geq \mathcal{F}(0)$, proving that with arbitrary control over $\theta$ pump enhancement gives sensitivities no worse than conventional SU(1,1) interferometry - and as we will demonstrate, usually much better in practice.

\begin{figure}[t]
\includegraphics[width=\columnwidth]{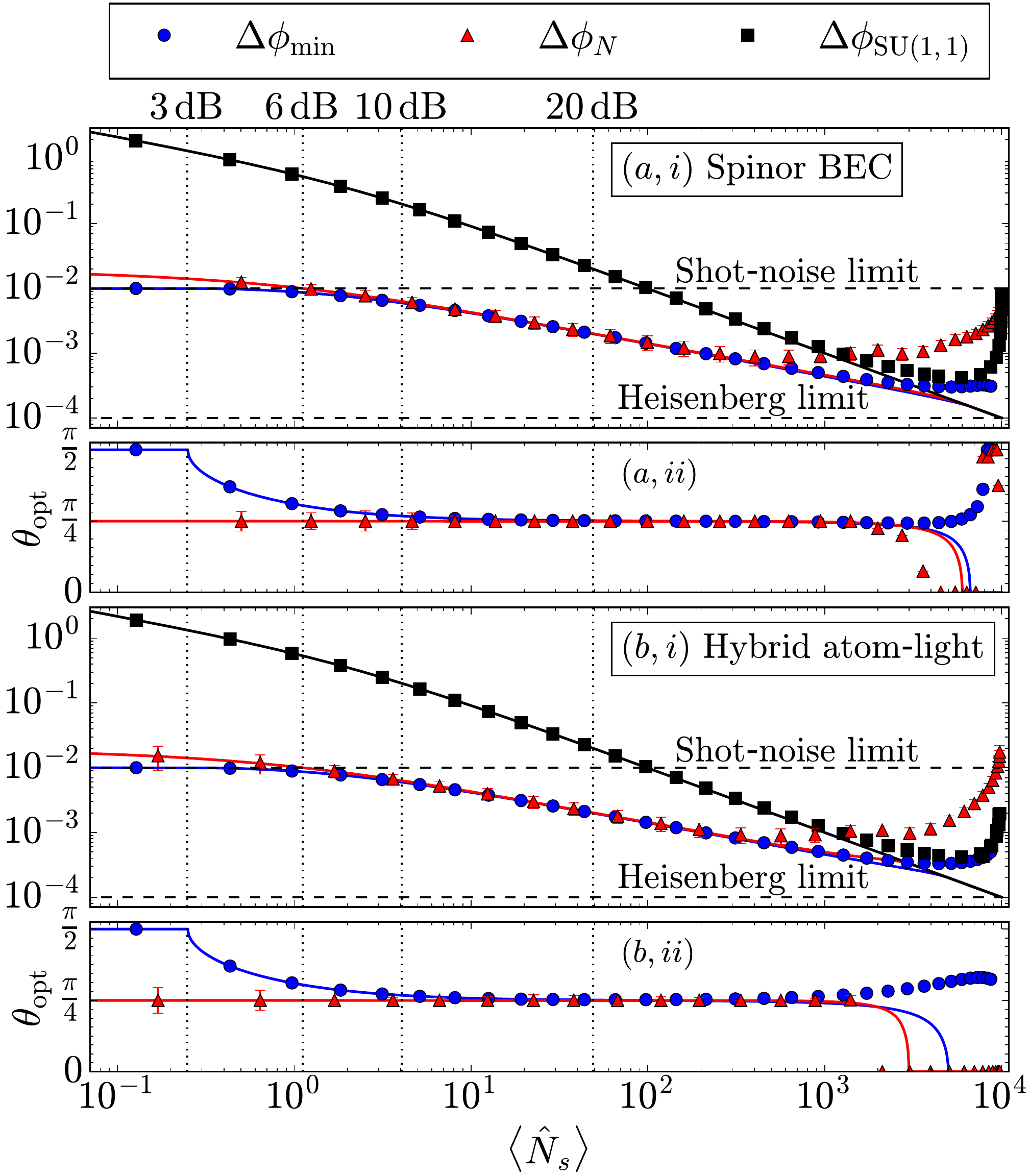}
\caption{Comparison of pumped-up and conventional SU(1,1) interferometry, engineered within (a) a spinor BEC and (b) a hybrid atom-light system (with $n_f = 1$). The total particle number is $\overline{N} = 10^4$. Sensitivities are plotted in (i), while optimal tritter (or beam splitter) angles $\theta_\text{opt}$ for pumped-up interferometry are shown in (ii). For our pumped-up schemes, $\Delta \phi_\text{min} = 1 / \sqrt{\mathcal{F}(\theta_\text{opt})}$ is the QCRB and $(\Delta \phi_N)^2 = \min_{\theta, \phi} \text{Var}(\hat{N}_s) / |\partial \langle \hat{N}_s \rangle/\partial \phi|^2$ gives the phase sensitivity for a number-sum measurement of the two side modes at the output; these are plotted for $\vartheta_\text{opt}$. $\Delta \phi_\text{SU(1,1)} = 1 / \sqrt{\mathcal{F}(0)}$ is the QCRB for conventional SU(1,1) interferometry, only saturated by a number-sum measurement of the side modes within the undepleted pump regime.
Solid lines are analytic curves obtained in the undepleted pump regime (accurate to all orders of $\overline{N}$ - see \cite{supplemental} for exact expressions), whereas markers are truncated Wigner simulations which include the effects of pump depletion \cite{DepletionFootnote}. The four vertical lines indicate the degree of squeezing associated with four values of $\langle \hat{N}_s \rangle$; these mark experimentally accessible regimes ranging from currently achievable (3~dB) to extremely challenging (20~dB). 
}
\label{fig_ideal}
\end{figure}

Maximizing Eq.~(\ref{QFI_tritter}) yields optimal parameters $\vartheta_\text{opt} = 3 \pi / 2$ and $\theta_\text{opt} = 0, \pi/2$, or $\left\{ \pi + 2 \text{csc}^{-1}[\mathcal{G}(\mathcal{N}_s, \vartheta) ] \right\}/4 + \mathcal{O}( 1 / \overline{N})$, and to leading order in $\overline{N}$ 
\begin{equation}
	\mathcal{F}(\theta_\text{opt}) = \begin{cases}
	 							\overline{N} + \tfrac{1}{2} \mathcal{N}_s^2, 			& \mathcal{N}_s < \tfrac{1}{4} \\
								\max \left\{ \frac{e^{2r}(1 + \coth r)}{8} \overline{N}, \mathcal{F}(0) \right\}, 	& \mathcal{N}_s \geq \tfrac{1}{4}	
	\end{cases}
\end{equation}
Therefore, pumped-up SU(1,1) interferometry has an achievable sensitivity at least as good as the shot-noise limit (with respect to total particle number $\overline{N}$), and any quantum enhancement improves the sensitivity \emph{beyond} this shot-noise limit. Conventional SU(1,1) is beneficial only when $\mathcal{N}_s$ is of the same order as $\overline{N}$, well outside both the undepleted pump regime and current experimental capabilities. Figure~\ref{fig_ideal}(a) graphically compares our pumped-up scheme with conventional SU(1,1) interferometry; this includes analytic undepleted pump expressions  and numerical truncated Wigner simulations \cite{Drummond:1993, Carter:1995, Olsen:2006} where $\hat{a}_0$ is treated as a quantum degree of freedom, thereby incorporating the effect of pump depletion \cite{supplemental}.

It was recently shown that the Loschmidt echo protocol saturates the QCRB \cite{Macri:2016}. In this protocol, the dynamics that evolved the initial state to the state with QFI $\mathcal{F}$ are reversed, and a measurement that projects the final state onto the initial state is made. For our scheme, this reversal corresponds to the second tritter and second spin-mixing step, followed by a measurement signal $\hat{\mathcal{S}}_\text{LE} = |\alpha_0, 0,0 \rangle\langle \alpha_0, 0,0|$. However, in practice superselection rules forbid measurements that project onto this initial pump coherent state; if instead we ignore the pump and choose a measurement signal $\hat{\mathcal{S}}_\text{LE}' = |0,0 \rangle\langle 0,0| = \sum_N |N,0,0\rangle \langle N,0,0|$ we obtain the \emph{suboptimal} sensitivity $\Delta \phi = 1 / \sqrt{\mathcal{F}(\theta) - \overline{N} \sin^4 \theta}$ \cite{supplemental}.

An operationally more convenient approach is to measure the number sum of the side modes at the outputs [as done in conventional SU(1,1) interferometry]. Although suboptimal, this phase measurement is more robust to inefficient detection than a Loschmidt echo \cite{Nolan_in_prep} and within the undepleted pump regime gives a phase sensitivity \cite{supplemental}
\begin{equation}
	\Delta \phi_{N} = \frac{\sqrt{\text{Var}(\hat{N}_s)}}{|\partial \langle \hat{N}_s \rangle / \partial \phi|} \Bigg|_{\phi = 0} = \frac{2|\text{csc}(2\theta)|}{\sqrt{ \eta(r) \overline{N}}} + \mathcal{O}\big(1/\overline{N}^{3/2}\big),
\end{equation}
where $\eta(r) \equiv \cosh(2r) - \sin (2\vartheta) \sinh(2r)$. Optimal parameters $\vartheta_\text{opt} = 3 \pi / 2$ and $\theta_\text{opt} = \pi / 4$ give minimum sensitivity $\Delta \phi_N \approx  2 \exp(-r) / \sqrt{\overline{N}}$. As confirmed in Fig.~\ref{fig_scheme}(b), this is never more than a factor of $2$ larger than the QCRB, and saturates this bound for $\mathcal{N}_s \gtrsim 2$.

\emph{Hybrid atom-light pumped-up SU(1,1) interferometry. ---} As shown in Fig.~\ref{fig_scheme}(c), the atomic and photonic pumps are mixed with their respective side modes via a variable angle two-mode beam splitter: $\hat{U}^a_\text{BS}(\theta) = \exp[-i \theta (e^{-i \vartheta} \hat{a}_0 \hat{a}_1^\dag + \textrm{H.c.})]$ and similarly for $\hat{U}^b_\text{BS}(\theta)$ [see Figs.~\ref{fig_scheme}(iv) and~\ref{fig_scheme}(v)]. The atomic modes are coupled via coherent light pulses commonly employed in atom interferometers \cite{Kasevich:1992}. For simplicity, we assume the atomic and photonic beam splitters have identical angle $\theta$ and phase $\vartheta$. As shown in Fig.~\ref{fig_ideal}(b), pumped-up SU(1,1) interferometry within a hybrid atom-light system has qualitative similarities to the spinor BEC case and therefore possesses all the same advantages over conventional SU(1,1) interferometry. One subtle difference is that the overall enhancement depends on both the total particle number $\overline{N}$ (atoms + photons) and the fraction of initial pump atoms to pump photons, $n_f$. Specifically, to leading order in $\overline{N}$, the maximum QFI and minimum phase sensitivity for a number-sum measurement are \cite{supplemental}
\begin{align}
	\mathcal{F}(\theta_\text{opt})	&= \begin{cases}
	 							\overline{N} - \mathcal{N}_s, 			& \mathcal{N}_s < \tfrac{1}{4} \\
								\max \left\{ \frac{\left[ \eta(r, n_f) \right]^2}{4[\eta(r, n_f) - 1]}\overline{N}, \mathcal{F}(0) \right\}, 	& \mathcal{N}_s \geq \tfrac{1}{4}	
	\end{cases} \\
	\Delta \phi_N		&= 2 / \sqrt{\eta(r, n_f) \overline{N}},
\end{align}
where $\eta(n_f) \equiv \cosh(2r) + [2 \sqrt{n_f} / (1+n_f)] \sinh(2r)$ and $\vartheta_\text{opt} = 3 \pi / 4$. For fixed $\overline{N}$, the optimal regime is $n_f = 1$, 
giving identical expressions to the spinor BEC case. More generally, there are likely to be considerably more photons than atoms ($n_f < 1$); since photons are ``cheap'' compared with atoms (in the sense that there are more severe particle-flux constraints on atoms than photons \cite{Szigeti:2012, Robins:2013}), a large absolute sensitivity could be obtained by increasing the number of pump photons (i.e. increasing $\overline{N}$) while simultaneously decreasing $n_f$ (therefore decreasing the per particle sensitivity), in the spirit of information recycling protocols \cite{Haine:2013, Szigeti:2014b, Tonekaboni:2015, Haine:2015, Haine:2015b, Haine:2016a}.

\begin{figure}%
\centering
\includegraphics[width=\columnwidth]{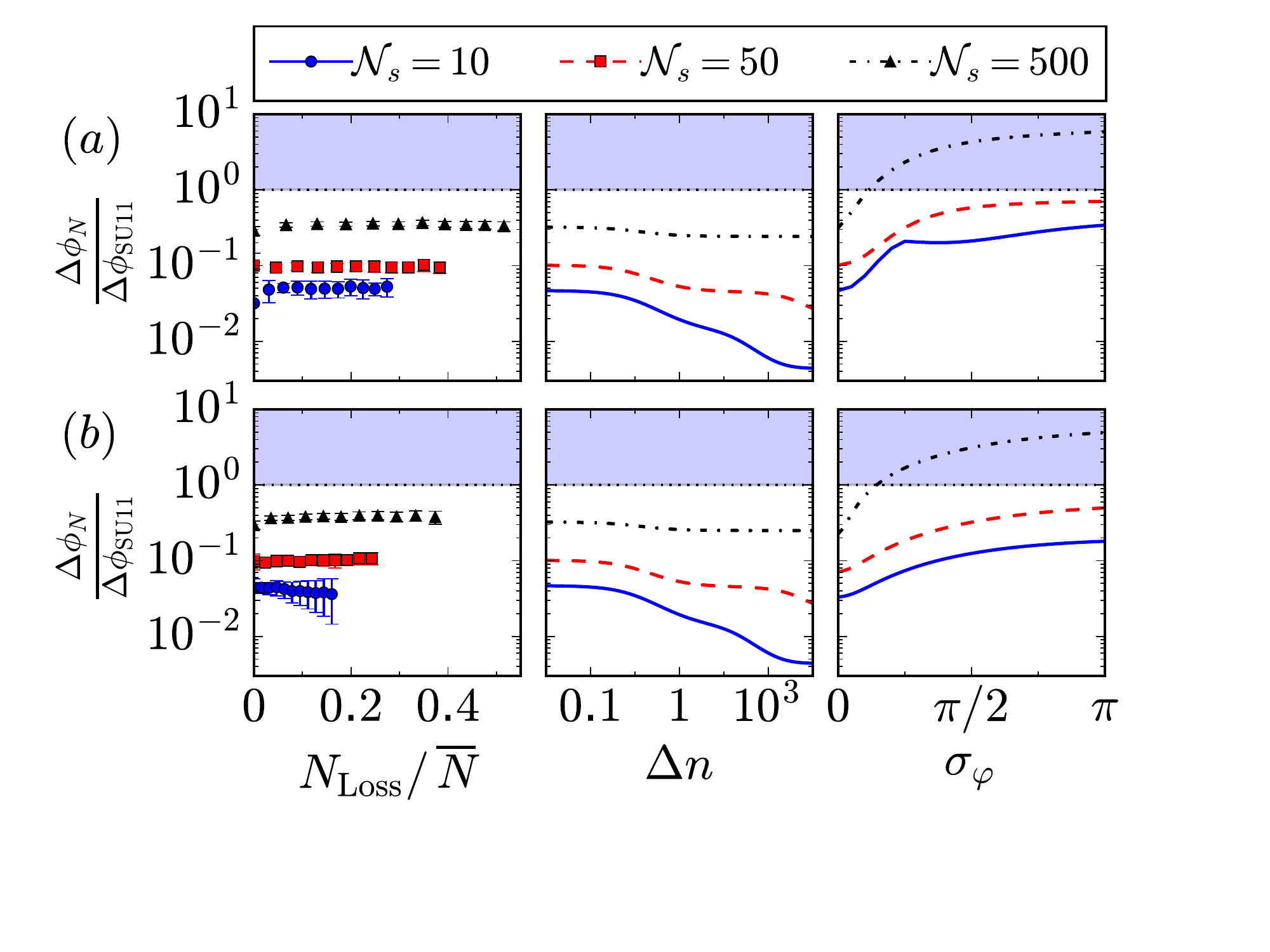}
\caption{Relative sensitivities of pumped-up SU(1,1) interferometry compared with conventional SU(1,1) interferometry in (a) a spinor BEC setup [top panels] and (b) a hybrid atom-light system (with $n_f=1$) [bottom panels]. The total particle number is $\overline{N} = 10^4$. All values plotted are at optimal $\phi$ and, for pumped-up schemes, optimal angle $\theta_\text{opt}$ and phase $\vartheta_\text{opt}$. These show the dependence on (left) the fraction of particles lost due to (a) two-body and (b) one-body losses, obtained via truncated Wigner simulations; (middle) imperfect particle detection with number resolution $\Delta n$, obtained from semianalytic calculations \cite{supplemental}; and (right) Gaussian phase-difference noise of variance $\sigma_\varphi^2$, obtained from analytic calculations with $\phi$ optimized numerically. Pumped-up SU(1,1) interferometry is superior to conventional SU(1,1) interferometry for those points/curves outside the shaded region. The side-mode populations $\mathcal{N}_s = 10$, $50$, and $500$ correspond to approximately $13.4$, $20$, and $30$~dB of squeezing, respectively. \emph{Absolute} sensitivities $\Delta \phi_N$ under losses are plotted in the Supplemental Material \cite{supplemental}.
}
\label{fig_losses}
\end{figure}

\emph{Effect of losses.---} Finally, we compare the performance of both pumped-up schemes to conventional SU(1,1) interferometry under the following three experimental sources of loss:

\emph{(i) Particle loss:} During spin-mixing dynamics of a spinor condensate particle loss is primarily caused by two-body recombination between atoms \cite{Tojo:2009, Lucke:2011, Gross:2011}, while for FWM within the hybrid atom-light system one-body particle losses are due to the spontaneous scattering of atoms and photons \cite{Moore:2000}.
Two-body losses during the spin-mixing dynamics are modeled with the master equation $\partial_t \hat{\rho} = -\tfrac{i}{\hbar}[\hat{H}_\text{SMD}, \hat{\rho}] + \sum_{i,j= 0, \pm} \gamma_{i,j} \mathcal{D}[\hat{a}_i \hat{a}_j] \hat{\rho}$, and one-body losses from the pumps during FWM with $\partial_t \hat{\rho} = -\tfrac{i}{\hbar}[\hat{H}_\text{FWM}, \hat{\rho}] + (\gamma_{a_0} \mathcal{D}[\hat{a}_0] + \gamma_{b_0} \mathcal{D}[\hat{b}_0])\hat{\rho}$, where $\mathcal{D}[\hat{L}]\hat{\rho} \equiv \hat{L} \hat{\rho} \hat{L}^{\dag} - \tfrac{1}{2} \{ \hat{L}^{\dagger} \hat{L}, \hat{\rho} \}$ and $\gamma_{i,j}$, $\gamma_{a_0}$, and $\gamma_{b_0}$ are loss rates. Since two-body loss is strongly number dependent, within the undepleted pump regime losses predominantly occur from the pump mode. Consequently, the precise value of loss rates involving collisions with $\hat{a}_\pm$ atoms relative to $\gamma_{0,0}$ is unimportant, so for simplicity we set $\gamma_{i,j} = \gamma$. We numerically solved these master equations and computed the phase sensitivity under the effect of these losses via the truncated Wigner simulation method \cite{supplemental}. As shown in the left panel in Fig.~\ref{fig_losses}, these types of particle loss affect pumped-up and conventional SU(1,1) interferometry similarly; consequently, our pumped-up approach maintains its considerable advantage.

\emph{(ii) Imperfect particle detection:} We model imperfect detection resolution as a Gaussian noise of variance $(\Delta n)^2$, which corresponds to an uncertainty $\Delta n$ in the particle number measured at the output. This technical noise increases the quantum noise on the signal, modifying the phase sensitivity: $(\Delta \phi_N)^2 = [\text{Var}(\hat{N}_s) + (\Delta n)^2] / (\partial_\phi \langle \hat{N}_s\rangle)^2$ \cite{Davis:2016}. In general, this modifies the optimal operating point; however, provided $\Delta n \lesssim \overline{N}$, the sensitivity of pumped-up SU(1,1) interferometry is \emph{independent} of imperfect particle detection \cite{supplemental}. This is a further advantage of pumped-up interferometry over conventional SU(1,1) interferometry. Furthermore, this robustness and superior performance is maintained for $\Delta n > \overline{N}$ [see the middle panel in Fig.~\ref{fig_losses}].

\emph{(iii) Phase difference noise:} In contrast to conventional SU(1,1) interferometry, our pumped-up schemes are sensitive to both the phase sum $\phi$ and phase \emph{difference} $\varphi$ between both arms of the interferometer. If an experiment cannot perfectly control $\varphi$ from shot-to-shot (e.g., energy shifts in spinor BECs due to the linear Zeeman effect), this degrades the sensitivity. We study the effect of this noise by assuming $\varphi$ is a Gaussian noise with variance $\sigma_\varphi^2$. As shown in the right panel in Fig.~\ref{fig_losses}, this degrades the sensitivity of pumped-up schemes compared with conventional SU(1,1) interferometry, particularly for larger values of quantum enhancement. Nevertheless, for the moderate levels of quantum enhancement achievable in practice, pumped-up SU(1,1) interferometry still surpasses conventional SU(1,1) interferometry between a factor of 2 and 10 - even for large $\sigma_\varphi$. Furthermore, the experimental results of \cite{Kruse:2016} suggest that noise due to $\varphi$ can be minimized in spinor BEC interferometers. 

\emph{Conclusions---} We have shown that pumped-up SU(1,1) interferometry considerably outperforms conventional SU(1,1) interferometry, even when typical experimental losses are included. Importantly, we illustrated the viability of pump enhancement in both spinor BECs and hybrid atom-light systems - which have both realized proof-of-principle conventional SU(1,1) interferometry, and are therefore capable of realizing our pumped-up schemes in the near term. Pumped-up SU(1,1) interferometry therefore pushes the advantages of conventional SU(1,1) interferometry into the regime of high absolute sensitivity, a necessary condition for useful quantum-enhanced devices.

We acknowledge useful discussions with Carlton Caves, Joel Corney, Daniel Linnemann, and Sam Nolan. Numerical simulations were performed using XMDS2 \cite{Dennis:2012} on the University of Queensland School of Mathematics and Physics computing cluster ``Obelix,'' with thanks to I. Mortimer for computing support. This project was supported by Australian Research Council (ARC) Project No.~DE130100575 and No.~DP140101763, and has received funding from the European Union's Horizon 2020 research and innovation programme under the Marie Sklodowska-Curie Grant Agreement No.~704672. S.~S.~S acknowledges the support of the ARC Centre of Excellence for Engineered Quantum Systems (Project No.~CE110001013).

\bibliography{tritter_PRL_bib.bib}

\begin{thebibliography}{58}%
\makeatletter
\providecommand \@ifxundefined [1]{%
 \@ifx{#1\undefined}
}%
\providecommand \@ifnum [1]{%
 \ifnum #1\expandafter \@firstoftwo
 \else \expandafter \@secondoftwo
 \fi
}%
\providecommand \@ifx [1]{%
 \ifx #1\expandafter \@firstoftwo
 \else \expandafter \@secondoftwo
 \fi
}%
\providecommand \natexlab [1]{#1}%
\providecommand \enquote  [1]{``#1''}%
\providecommand \bibnamefont  [1]{#1}%
\providecommand \bibfnamefont [1]{#1}%
\providecommand \citenamefont [1]{#1}%
\providecommand \href@noop [0]{\@secondoftwo}%
\providecommand \href [0]{\begingroup \@sanitize@url \@href}%
\providecommand \@href[1]{\@@startlink{#1}\@@href}%
\providecommand \@@href[1]{\endgroup#1\@@endlink}%
\providecommand \@sanitize@url [0]{\catcode `\\12\catcode `\$12\catcode
  `\&12\catcode `\#12\catcode `\^12\catcode `\_12\catcode `\%12\relax}%
\providecommand \@@startlink[1]{}%
\providecommand \@@endlink[0]{}%
\providecommand \url  [0]{\begingroup\@sanitize@url \@url }%
\providecommand \@url [1]{\endgroup\@href {#1}{\urlprefix }}%
\providecommand \urlprefix  [0]{URL }%
\providecommand \Eprint [0]{\href }%
\providecommand \doibase [0]{http://dx.doi.org/}%
\providecommand \selectlanguage [0]{\@gobble}%
\providecommand \bibinfo  [0]{\@secondoftwo}%
\providecommand \bibfield  [0]{\@secondoftwo}%
\providecommand \translation [1]{[#1]}%
\providecommand \BibitemOpen [0]{}%
\providecommand \bibitemStop [0]{}%
\providecommand \bibitemNoStop [0]{.\EOS\space}%
\providecommand \EOS [0]{\spacefactor3000\relax}%
\providecommand \BibitemShut  [1]{\csname bibitem#1\endcsname}%
\let\auto@bib@innerbib\@empty
\bibitem [{\citenamefont {Caves}(1981)}]{Caves:1981}%
  \BibitemOpen
  \bibfield  {author} {\bibinfo {author} {\bibfnamefont {C.~M.}\ \bibnamefont
  {Caves}},\ }\href {\doibase 10.1103/PhysRevD.23.1693} {\bibfield  {journal}
  {\bibinfo  {journal} {Phys. Rev. D}\ }\textbf {\bibinfo {volume} {23}},\
  \bibinfo {pages} {1693} (\bibinfo {year} {1981})}\BibitemShut {NoStop}%
\bibitem [{\citenamefont {T\'oth}\ and\ \citenamefont
  {Apellaniz}(2014)}]{Toth:2014}%
  \BibitemOpen
  \bibfield  {author} {\bibinfo {author} {\bibfnamefont {G.}~\bibnamefont
  {T\'oth}}\ and\ \bibinfo {author} {\bibfnamefont {I.}~\bibnamefont
  {Apellaniz}},\ }\href {http://stacks.iop.org/1751-8121/47/i=42/a=424006}
  {\bibfield  {journal} {\bibinfo  {journal} {Journal of Physics A:
  Mathematical and Theoretical}\ }\textbf {\bibinfo {volume} {47}},\ \bibinfo
  {pages} {424006} (\bibinfo {year} {2014})}\BibitemShut {NoStop}%
\bibitem [{\citenamefont {{The LIGO Scientific
  Collaboration}}(2011)}]{LIGO:2011}%
  \BibitemOpen
  \bibfield  {author} {\bibinfo {author} {\bibnamefont {{The LIGO Scientific
  Collaboration}}},\ }\href {http://dx.doi.org/10.1038/nphys2083} {\bibfield
  {journal} {\bibinfo  {journal} {Nat. Phys.}\ }\textbf {\bibinfo {volume}
  {7}},\ \bibinfo {pages} {962} (\bibinfo {year} {2011})}\BibitemShut {NoStop}%
\bibitem [{\citenamefont {{The LIGO Scientific
  Collaboration}}(2013)}]{LIGO:2013}%
  \BibitemOpen
  \bibfield  {author} {\bibinfo {author} {\bibnamefont {{The LIGO Scientific
  Collaboration}}},\ }\href {http://dx.doi.org/10.1038/nphoton.2013.177}
  {\bibfield  {journal} {\bibinfo  {journal} {Nat. Photon.}\ }\textbf {\bibinfo
  {volume} {7}},\ \bibinfo {pages} {613} (\bibinfo {year} {2013})}\BibitemShut
  {NoStop}%
\bibitem [{\citenamefont {Demkowicz-Dobrzanski}\ \emph
  {et~al.}(2012)\citenamefont {Demkowicz-Dobrzanski}, \citenamefont
  {Kolodynski},\ and\ \citenamefont {Guta}}]{Demkowicz-Dobrzanski:2012}%
  \BibitemOpen
  \bibfield  {author} {\bibinfo {author} {\bibfnamefont {R.}~\bibnamefont
  {Demkowicz-Dobrzanski}}, \bibinfo {author} {\bibfnamefont {J.}~\bibnamefont
  {Kolodynski}}, \ and\ \bibinfo {author} {\bibfnamefont {M.}~\bibnamefont
  {Guta}},\ }\href {http://dx.doi.org/10.1038/ncomms2067} {\bibfield  {journal}
  {\bibinfo  {journal} {Nat. Commun.}\ }\textbf {\bibinfo {volume} {3}},\
  \bibinfo {pages} {1063} (\bibinfo {year} {2012})}\BibitemShut {NoStop}%
\bibitem [{\citenamefont {Yurke}\ \emph {et~al.}(1986)\citenamefont {Yurke},
  \citenamefont {McCall},\ and\ \citenamefont {Klauder}}]{Yurke:1986}%
  \BibitemOpen
  \bibfield  {author} {\bibinfo {author} {\bibfnamefont {B.}~\bibnamefont
  {Yurke}}, \bibinfo {author} {\bibfnamefont {S.~L.}\ \bibnamefont {McCall}}, \
  and\ \bibinfo {author} {\bibfnamefont {J.~R.}\ \bibnamefont {Klauder}},\
  }\href {\doibase 10.1103/PhysRevA.33.4033} {\bibfield  {journal} {\bibinfo
  {journal} {Phys. Rev. A}\ }\textbf {\bibinfo {volume} {33}},\ \bibinfo
  {pages} {4033} (\bibinfo {year} {1986})}\BibitemShut {NoStop}%
\bibitem [{\citenamefont {Leonhardt}(1994)}]{Leonhardt:1994}%
  \BibitemOpen
  \bibfield  {author} {\bibinfo {author} {\bibfnamefont {U.}~\bibnamefont
  {Leonhardt}},\ }\href {\doibase 10.1103/PhysRevA.49.1231} {\bibfield
  {journal} {\bibinfo  {journal} {Phys. Rev. A}\ }\textbf {\bibinfo {volume}
  {49}},\ \bibinfo {pages} {1231} (\bibinfo {year} {1994})}\BibitemShut
  {NoStop}%
\bibitem [{\citenamefont {Marino}\ \emph {et~al.}(2012)\citenamefont {Marino},
  \citenamefont {Corzo~Trejo},\ and\ \citenamefont {Lett}}]{Marino:2012}%
  \BibitemOpen
  \bibfield  {author} {\bibinfo {author} {\bibfnamefont {A.~M.}\ \bibnamefont
  {Marino}}, \bibinfo {author} {\bibfnamefont {N.~V.}\ \bibnamefont
  {Corzo~Trejo}}, \ and\ \bibinfo {author} {\bibfnamefont {P.~D.}\ \bibnamefont
  {Lett}},\ }\href {\doibase 10.1103/PhysRevA.86.023844} {\bibfield  {journal}
  {\bibinfo  {journal} {Phys. Rev. A}\ }\textbf {\bibinfo {volume} {86}},\
  \bibinfo {pages} {023844} (\bibinfo {year} {2012})}\BibitemShut {NoStop}%
\bibitem [{\citenamefont {Ou}(2012)}]{Ou:2012}%
  \BibitemOpen
  \bibfield  {author} {\bibinfo {author} {\bibfnamefont {Z.~Y.}\ \bibnamefont
  {Ou}},\ }\href {\doibase 10.1103/PhysRevA.85.023815} {\bibfield  {journal}
  {\bibinfo  {journal} {Phys. Rev. A}\ }\textbf {\bibinfo {volume} {85}},\
  \bibinfo {pages} {023815} (\bibinfo {year} {2012})}\BibitemShut {NoStop}%
\bibitem [{\citenamefont {Gabbrielli}\ \emph {et~al.}(2015)\citenamefont
  {Gabbrielli}, \citenamefont {Pezz\`e},\ and\ \citenamefont
  {Smerzi}}]{Gabbrielli:2015}%
  \BibitemOpen
  \bibfield  {author} {\bibinfo {author} {\bibfnamefont {M.}~\bibnamefont
  {Gabbrielli}}, \bibinfo {author} {\bibfnamefont {L.}~\bibnamefont {Pezz\`e}},
  \ and\ \bibinfo {author} {\bibfnamefont {A.}~\bibnamefont {Smerzi}},\ }\href
  {\doibase 10.1103/PhysRevLett.115.163002} {\bibfield  {journal} {\bibinfo
  {journal} {Phys. Rev. Lett.}\ }\textbf {\bibinfo {volume} {115}},\ \bibinfo
  {pages} {163002} (\bibinfo {year} {2015})}\BibitemShut {NoStop}%
\bibitem [{\citenamefont {Chen}\ \emph {et~al.}(2016)\citenamefont {Chen},
  \citenamefont {Yuan}, \citenamefont {Ma}, \citenamefont {Li}, \citenamefont
  {Chen}, \citenamefont {Ou},\ and\ \citenamefont {Zhang}}]{Chen:2016}%
  \BibitemOpen
  \bibfield  {author} {\bibinfo {author} {\bibfnamefont {Z.-D.}\ \bibnamefont
  {Chen}}, \bibinfo {author} {\bibfnamefont {C.-H.}\ \bibnamefont {Yuan}},
  \bibinfo {author} {\bibfnamefont {H.-M.}\ \bibnamefont {Ma}}, \bibinfo
  {author} {\bibfnamefont {D.}~\bibnamefont {Li}}, \bibinfo {author}
  {\bibfnamefont {L.~Q.}\ \bibnamefont {Chen}}, \bibinfo {author}
  {\bibfnamefont {Z.~Y.}\ \bibnamefont {Ou}}, \ and\ \bibinfo {author}
  {\bibfnamefont {W.}~\bibnamefont {Zhang}},\ }\bibfield  {booktitle} {\emph
  {\bibinfo {booktitle} {Optics Express}},\ }\href {\doibase
  10.1364/OE.24.017766} {\bibfield  {journal} {\bibinfo  {journal} {Optics
  Express}\ }\textbf {\bibinfo {volume} {24}},\ \bibinfo {pages} {17766}
  (\bibinfo {year} {2016})}\BibitemShut {NoStop}%
\bibitem [{\citenamefont {Gong}\ \emph {et~al.}(2016)\citenamefont {Gong},
  \citenamefont {Li}, \citenamefont {Yuan}, \citenamefont {Ou},\ and\
  \citenamefont {Zhang}}]{Gong:2016}%
  \BibitemOpen
  \bibfield  {author} {\bibinfo {author} {\bibfnamefont {Q.-K.}\ \bibnamefont
  {Gong}}, \bibinfo {author} {\bibfnamefont {D.}~\bibnamefont {Li}}, \bibinfo
  {author} {\bibfnamefont {C.-H.}\ \bibnamefont {Yuan}}, \bibinfo {author}
  {\bibfnamefont {Z.~Y.}\ \bibnamefont {Ou}}, \ and\ \bibinfo {author}
  {\bibfnamefont {W.}~\bibnamefont {Zhang}},\ }\href@noop {} {\bibfield
  {journal} {\bibinfo  {journal} {arXiv:1608.00389v2}\ } (\bibinfo {year}
  {2016})}\BibitemShut {NoStop}%
\bibitem [{\citenamefont {Xin}\ \emph {et~al.}(2016)\citenamefont {Xin},
  \citenamefont {Wang},\ and\ \citenamefont {Jing}}]{Xin:2016}%
  \BibitemOpen
  \bibfield  {author} {\bibinfo {author} {\bibfnamefont {J.}~\bibnamefont
  {Xin}}, \bibinfo {author} {\bibfnamefont {H.}~\bibnamefont {Wang}}, \ and\
  \bibinfo {author} {\bibfnamefont {J.}~\bibnamefont {Jing}},\ }\href {\doibase
  http://dx.doi.org/10.1063/1.4960585} {\bibfield  {journal} {\bibinfo
  {journal} {Applied Physics Letters}\ }\textbf {\bibinfo {volume} {109}},\
  \bibinfo {eid} {051107} (\bibinfo {year} {2016}),\
  http://dx.doi.org/10.1063/1.4960585}\BibitemShut {NoStop}%
\bibitem [{\citenamefont {Jing}\ \emph {et~al.}(2011)\citenamefont {Jing},
  \citenamefont {Liu}, \citenamefont {Zhou}, \citenamefont {Ou},\ and\
  \citenamefont {Zhang}}]{Jing:2011}%
  \BibitemOpen
  \bibfield  {author} {\bibinfo {author} {\bibfnamefont {J.}~\bibnamefont
  {Jing}}, \bibinfo {author} {\bibfnamefont {C.}~\bibnamefont {Liu}}, \bibinfo
  {author} {\bibfnamefont {Z.}~\bibnamefont {Zhou}}, \bibinfo {author}
  {\bibfnamefont {Z.~Y.}\ \bibnamefont {Ou}}, \ and\ \bibinfo {author}
  {\bibfnamefont {W.}~\bibnamefont {Zhang}},\ }\href {\doibase
  http://dx.doi.org/10.1063/1.3606549} {\bibfield  {journal} {\bibinfo
  {journal} {Applied Physics Letters}\ }\textbf {\bibinfo {volume} {99}},\
  \bibinfo {eid} {011110} (\bibinfo {year} {2011}),\
  http://dx.doi.org/10.1063/1.3606549}\BibitemShut {NoStop}%
\bibitem [{\citenamefont {Hudelist}\ \emph {et~al.}(2014)\citenamefont
  {Hudelist}, \citenamefont {Kong}, \citenamefont {Liu}, \citenamefont {Jing},
  \citenamefont {Ou},\ and\ \citenamefont {Zhang}}]{Hudelist:2014}%
  \BibitemOpen
  \bibfield  {author} {\bibinfo {author} {\bibfnamefont {F.}~\bibnamefont
  {Hudelist}}, \bibinfo {author} {\bibfnamefont {J.}~\bibnamefont {Kong}},
  \bibinfo {author} {\bibfnamefont {C.}~\bibnamefont {Liu}}, \bibinfo {author}
  {\bibfnamefont {J.}~\bibnamefont {Jing}}, \bibinfo {author} {\bibfnamefont
  {Z.~Y.}\ \bibnamefont {Ou}}, \ and\ \bibinfo {author} {\bibfnamefont
  {W.}~\bibnamefont {Zhang}},\ }\href {http://dx.doi.org/10.1038/ncomms4049}
  {\bibfield  {journal} {\bibinfo  {journal} {Nature Communications}\ }\textbf
  {\bibinfo {volume} {5}},\ \bibinfo {pages} {3049 EP } (\bibinfo {year}
  {2014})}\BibitemShut {NoStop}%
\bibitem [{\citenamefont {Chen}\ \emph {et~al.}(2015)\citenamefont {Chen},
  \citenamefont {Qiu}, \citenamefont {Chen}, \citenamefont {Guo}, \citenamefont
  {Chen}, \citenamefont {Ou},\ and\ \citenamefont {Zhang}}]{Chen:2015b}%
  \BibitemOpen
  \bibfield  {author} {\bibinfo {author} {\bibfnamefont {B.}~\bibnamefont
  {Chen}}, \bibinfo {author} {\bibfnamefont {C.}~\bibnamefont {Qiu}}, \bibinfo
  {author} {\bibfnamefont {S.}~\bibnamefont {Chen}}, \bibinfo {author}
  {\bibfnamefont {J.}~\bibnamefont {Guo}}, \bibinfo {author} {\bibfnamefont
  {L.~Q.}\ \bibnamefont {Chen}}, \bibinfo {author} {\bibfnamefont {Z.~Y.}\
  \bibnamefont {Ou}}, \ and\ \bibinfo {author} {\bibfnamefont {W.}~\bibnamefont
  {Zhang}},\ }\href {\doibase 10.1103/PhysRevLett.115.043602} {\bibfield
  {journal} {\bibinfo  {journal} {Phys. Rev. Lett.}\ }\textbf {\bibinfo
  {volume} {115}},\ \bibinfo {pages} {043602} (\bibinfo {year}
  {2015})}\BibitemShut {NoStop}%
\bibitem [{\citenamefont {Gross}\ \emph {et~al.}(2010)\citenamefont {Gross},
  \citenamefont {Zibold}, \citenamefont {Nicklas}, \citenamefont {Est{\`e}ve},\
  and\ \citenamefont {Oberthaler}}]{Gross:2010}%
  \BibitemOpen
  \bibfield  {author} {\bibinfo {author} {\bibfnamefont {C.}~\bibnamefont
  {Gross}}, \bibinfo {author} {\bibfnamefont {T.}~\bibnamefont {Zibold}},
  \bibinfo {author} {\bibfnamefont {E.}~\bibnamefont {Nicklas}}, \bibinfo
  {author} {\bibfnamefont {J.}~\bibnamefont {Est{\`e}ve}}, \ and\ \bibinfo
  {author} {\bibfnamefont {M.~K.}\ \bibnamefont {Oberthaler}},\ }\href
  {http://dx.doi.org/10.1038/nature08919} {\bibfield  {journal} {\bibinfo
  {journal} {Nature}\ }\textbf {\bibinfo {volume} {464}},\ \bibinfo {pages}
  {1165} (\bibinfo {year} {2010})}\BibitemShut {NoStop}%
\bibitem [{\citenamefont {Peise}\ \emph {et~al.}(2015)\citenamefont {Peise},
  \citenamefont {L{\"u}cke}, \citenamefont {Pezz{\'e}}, \citenamefont
  {Deuretzbacher}, \citenamefont {Ertmer}, \citenamefont {Arlt}, \citenamefont
  {Smerzi}, \citenamefont {Santos},\ and\ \citenamefont {Klempt}}]{Peise:2015}%
  \BibitemOpen
  \bibfield  {author} {\bibinfo {author} {\bibfnamefont {J.}~\bibnamefont
  {Peise}}, \bibinfo {author} {\bibfnamefont {B.}~\bibnamefont {L{\"u}cke}},
  \bibinfo {author} {\bibfnamefont {L.}~\bibnamefont {Pezz{\'e}}}, \bibinfo
  {author} {\bibfnamefont {F.}~\bibnamefont {Deuretzbacher}}, \bibinfo {author}
  {\bibfnamefont {W.}~\bibnamefont {Ertmer}}, \bibinfo {author} {\bibfnamefont
  {J.}~\bibnamefont {Arlt}}, \bibinfo {author} {\bibfnamefont {A.}~\bibnamefont
  {Smerzi}}, \bibinfo {author} {\bibfnamefont {L.}~\bibnamefont {Santos}}, \
  and\ \bibinfo {author} {\bibfnamefont {C.}~\bibnamefont {Klempt}},\ }\href
  {http://dx.doi.org/10.1038/ncomms7811} {\bibfield  {journal} {\bibinfo
  {journal} {Nature Communications}\ }\textbf {\bibinfo {volume} {6}},\
  \bibinfo {pages} {6811 EP } (\bibinfo {year} {2015})}\BibitemShut {NoStop}%
\bibitem [{\citenamefont {Linnemann}\ \emph {et~al.}(2016)\citenamefont
  {Linnemann}, \citenamefont {Strobel}, \citenamefont {Muessel}, \citenamefont
  {Schulz}, \citenamefont {Lewis-Swan}, \citenamefont {Kheruntsyan},\ and\
  \citenamefont {Oberthaler}}]{Linnemann:2016}%
  \BibitemOpen
  \bibfield  {author} {\bibinfo {author} {\bibfnamefont {D.}~\bibnamefont
  {Linnemann}}, \bibinfo {author} {\bibfnamefont {H.}~\bibnamefont {Strobel}},
  \bibinfo {author} {\bibfnamefont {W.}~\bibnamefont {Muessel}}, \bibinfo
  {author} {\bibfnamefont {J.}~\bibnamefont {Schulz}}, \bibinfo {author}
  {\bibfnamefont {R.~J.}\ \bibnamefont {Lewis-Swan}}, \bibinfo {author}
  {\bibfnamefont {K.~V.}\ \bibnamefont {Kheruntsyan}}, \ and\ \bibinfo {author}
  {\bibfnamefont {M.~K.}\ \bibnamefont {Oberthaler}},\ }\href {\doibase
  10.1103/PhysRevLett.117.013001} {\bibfield  {journal} {\bibinfo  {journal}
  {Phys. Rev. Lett.}\ }\textbf {\bibinfo {volume} {117}},\ \bibinfo {pages}
  {013001} (\bibinfo {year} {2016})}\BibitemShut {NoStop}%
\bibitem [{\citenamefont {Loudon}\ and\ \citenamefont
  {Knight}(1987)}]{Loudon:1987}%
  \BibitemOpen
  \bibfield  {author} {\bibinfo {author} {\bibfnamefont {R.}~\bibnamefont
  {Loudon}}\ and\ \bibinfo {author} {\bibfnamefont {P.~L.}\ \bibnamefont
  {Knight}},\ }\bibfield  {booktitle} {\emph {\bibinfo {booktitle} {Journal of
  Modern Optics}},\ }\href {\doibase 10.1080/09500348714550721} {\bibfield
  {journal} {\bibinfo  {journal} {Journal of Modern Optics}\ }\textbf {\bibinfo
  {volume} {34}},\ \bibinfo {pages} {709} (\bibinfo {year} {1987})}\BibitemShut
  {NoStop}%
\bibitem [{\citenamefont {Law}\ \emph {et~al.}(1998)\citenamefont {Law},
  \citenamefont {Pu},\ and\ \citenamefont {Bigelow}}]{Law:1998}%
  \BibitemOpen
  \bibfield  {author} {\bibinfo {author} {\bibfnamefont {C.~K.}\ \bibnamefont
  {Law}}, \bibinfo {author} {\bibfnamefont {H.}~\bibnamefont {Pu}}, \ and\
  \bibinfo {author} {\bibfnamefont {N.~P.}\ \bibnamefont {Bigelow}},\ }\href
  {\doibase 10.1103/PhysRevLett.81.5257} {\bibfield  {journal} {\bibinfo
  {journal} {Phys. Rev. Lett.}\ }\textbf {\bibinfo {volume} {81}},\ \bibinfo
  {pages} {5257} (\bibinfo {year} {1998})}\BibitemShut {NoStop}%
\bibitem [{\citenamefont {Lewis-Swan}\ and\ \citenamefont
  {Kheruntsyan}(2013)}]{Lewis-Swan:2013}%
  \BibitemOpen
  \bibfield  {author} {\bibinfo {author} {\bibfnamefont {R.~J.}\ \bibnamefont
  {Lewis-Swan}}\ and\ \bibinfo {author} {\bibfnamefont {K.~V.}\ \bibnamefont
  {Kheruntsyan}},\ }\href {\doibase 10.1103/PhysRevA.87.063635} {\bibfield
  {journal} {\bibinfo  {journal} {Phys. Rev. A}\ }\textbf {\bibinfo {volume}
  {87}},\ \bibinfo {pages} {063635} (\bibinfo {year} {2013})}\BibitemShut
  {NoStop}%
\bibitem [{Zee()}]{Zeeman_shift}%
  \BibitemOpen
  \href@noop {} {}\bibinfo {howpublished} {In undepleted pump regime $q(t) =
  \kappa(\langle \hat{N}_0 \rangle - 1/2)$ suffices.}\BibitemShut {Stop}%
\bibitem [{\citenamefont {Moore}\ and\ \citenamefont
  {Meystre}(2000)}]{Moore:2000}%
  \BibitemOpen
  \bibfield  {author} {\bibinfo {author} {\bibfnamefont {M.~G.}\ \bibnamefont
  {Moore}}\ and\ \bibinfo {author} {\bibfnamefont {P.}~\bibnamefont
  {Meystre}},\ }\href {\doibase 10.1103/PhysRevLett.85.5026} {\bibfield
  {journal} {\bibinfo  {journal} {Phys. Rev. Lett.}\ }\textbf {\bibinfo
  {volume} {85}},\ \bibinfo {pages} {5026} (\bibinfo {year}
  {2000})}\BibitemShut {NoStop}%
\bibitem [{\citenamefont {Hammerer}\ \emph {et~al.}(2010)\citenamefont
  {Hammerer}, \citenamefont {S\o{}rensen},\ and\ \citenamefont
  {Polzik}}]{Hammerer:2010}%
  \BibitemOpen
  \bibfield  {author} {\bibinfo {author} {\bibfnamefont {K.}~\bibnamefont
  {Hammerer}}, \bibinfo {author} {\bibfnamefont {A.~S.}\ \bibnamefont
  {S\o{}rensen}}, \ and\ \bibinfo {author} {\bibfnamefont {E.~S.}\ \bibnamefont
  {Polzik}},\ }\href {\doibase 10.1103/RevModPhys.82.1041} {\bibfield
  {journal} {\bibinfo  {journal} {Rev. Mod. Phys.}\ }\textbf {\bibinfo {volume}
  {82}},\ \bibinfo {pages} {1041} (\bibinfo {year} {2010})}\BibitemShut
  {NoStop}%
\bibitem [{\citenamefont {Haine}(2013)}]{Haine:2013}%
  \BibitemOpen
  \bibfield  {author} {\bibinfo {author} {\bibfnamefont {S.~A.}\ \bibnamefont
  {Haine}},\ }\href {\doibase 10.1103/PhysRevLett.110.053002} {\bibfield
  {journal} {\bibinfo  {journal} {Phys. Rev. Lett.}\ }\textbf {\bibinfo
  {volume} {110}},\ \bibinfo {pages} {053002} (\bibinfo {year}
  {2013})}\BibitemShut {NoStop}%
\bibitem [{\citenamefont {Haine}\ and\ \citenamefont
  {Lau}(2016)}]{Haine:2016a}%
  \BibitemOpen
  \bibfield  {author} {\bibinfo {author} {\bibfnamefont {S.~A.}\ \bibnamefont
  {Haine}}\ and\ \bibinfo {author} {\bibfnamefont {W.~Y.~S.}\ \bibnamefont
  {Lau}},\ }\href {\doibase 10.1103/PhysRevA.93.023607} {\bibfield  {journal}
  {\bibinfo  {journal} {Phys. Rev. A}\ }\textbf {\bibinfo {volume} {93}},\
  \bibinfo {pages} {023607} (\bibinfo {year} {2016})}\BibitemShut {NoStop}%
\bibitem [{\citenamefont {Kruse}\ \emph {et~al.}(2016)\citenamefont {Kruse},
  \citenamefont {Lange}, \citenamefont {Peise}, \citenamefont {L\"ucke},
  \citenamefont {Pezz\`e}, \citenamefont {Arlt}, \citenamefont {Ertmer},
  \citenamefont {Lisdat}, \citenamefont {Santos}, \citenamefont {Smerzi},\ and\
  \citenamefont {Klempt}}]{Kruse:2016}%
  \BibitemOpen
  \bibfield  {author} {\bibinfo {author} {\bibfnamefont {I.}~\bibnamefont
  {Kruse}}, \bibinfo {author} {\bibfnamefont {K.}~\bibnamefont {Lange}},
  \bibinfo {author} {\bibfnamefont {J.}~\bibnamefont {Peise}}, \bibinfo
  {author} {\bibfnamefont {B.}~\bibnamefont {L\"ucke}}, \bibinfo {author}
  {\bibfnamefont {L.}~\bibnamefont {Pezz\`e}}, \bibinfo {author} {\bibfnamefont
  {J.}~\bibnamefont {Arlt}}, \bibinfo {author} {\bibfnamefont {W.}~\bibnamefont
  {Ertmer}}, \bibinfo {author} {\bibfnamefont {C.}~\bibnamefont {Lisdat}},
  \bibinfo {author} {\bibfnamefont {L.}~\bibnamefont {Santos}}, \bibinfo
  {author} {\bibfnamefont {A.}~\bibnamefont {Smerzi}}, \ and\ \bibinfo {author}
  {\bibfnamefont {C.}~\bibnamefont {Klempt}},\ }\href {\doibase
  10.1103/PhysRevLett.117.143004} {\bibfield  {journal} {\bibinfo  {journal}
  {Phys. Rev. Lett.}\ }\textbf {\bibinfo {volume} {117}},\ \bibinfo {pages}
  {143004} (\bibinfo {year} {2016})}\BibitemShut {NoStop}%
\bibitem [{\citenamefont {Braunstein}\ and\ \citenamefont
  {Caves}(1994)}]{Braunstein:1994}%
  \BibitemOpen
  \bibfield  {author} {\bibinfo {author} {\bibfnamefont {S.~L.}\ \bibnamefont
  {Braunstein}}\ and\ \bibinfo {author} {\bibfnamefont {C.~M.}\ \bibnamefont
  {Caves}},\ }\href {\doibase 10.1103/PhysRevLett.72.3439} {\bibfield
  {journal} {\bibinfo  {journal} {Phys. Rev. Lett.}\ }\textbf {\bibinfo
  {volume} {72}},\ \bibinfo {pages} {3439} (\bibinfo {year}
  {1994})}\BibitemShut {NoStop}%
\bibitem [{\citenamefont {Demkowicz-Dobrza{\'n}ski}\ \emph
  {et~al.}(2015)\citenamefont {Demkowicz-Dobrza{\'n}ski}, \citenamefont
  {Jarzyna}, \citenamefont {Ko{\l}ody{\'n}ski},\ and\ \citenamefont
  {Wolf}}]{Demkowicz-Dobrzanski:2015}%
  \BibitemOpen
  \bibfield  {author} {\bibinfo {author} {\bibfnamefont {R.}~\bibnamefont
  {Demkowicz-Dobrza{\'n}ski}}, \bibinfo {author} {\bibfnamefont
  {M.}~\bibnamefont {Jarzyna}}, \bibinfo {author} {\bibfnamefont
  {J.}~\bibnamefont {Ko{\l}ody{\'n}ski}}, \ and\ \bibinfo {author}
  {\bibfnamefont {E.}~\bibnamefont {Wolf}},\ }\enquote {\bibinfo {title}
  {Chapter four - quantum limits in optical interferometry},}\ in\ \href
  {\doibase http://dx.doi.org/10.1016/bs.po.2015.02.003} {\emph {\bibinfo
  {booktitle} {Progress in Optics}}},\ Vol.\ \bibinfo {volume} {Volume 60}\
  (\bibinfo  {publisher} {Elsevier},\ \bibinfo {year} {2015})\ pp.\ \bibinfo
  {pages} {345--435}\BibitemShut {NoStop}%
\bibitem [{sup()}]{supplemental}%
  \BibitemOpen
  \href@noop {} {}\bibinfo {howpublished} {See supplemental material, which
  includes Refs \cite{Macri:2016, Steel:1998, Blakie:2008, Polkovnikov:2010,
  Opanchuk:2013, Walls:2008,
  Gardiner:2004b,Sinatra:2002,Opanchuk:2012,Olsen:2009}, for detailed
  calculations of the quantum Fisher information and phase sensitivity of a
  number-sum measurement, and a brief description of our truncated Wigner
  simulations.}\BibitemShut {Stop}%
\bibitem [{\citenamefont {Macr\`{\i}}\ \emph {et~al.}(2016)\citenamefont
  {Macr\`{\i}}, \citenamefont {Smerzi},\ and\ \citenamefont
  {Pezz\`e}}]{Macri:2016}%
  \BibitemOpen
  \bibfield  {author} {\bibinfo {author} {\bibfnamefont {T.}~\bibnamefont
  {Macr\`{\i}}}, \bibinfo {author} {\bibfnamefont {A.}~\bibnamefont {Smerzi}},
  \ and\ \bibinfo {author} {\bibfnamefont {L.}~\bibnamefont {Pezz\`e}},\ }\href
  {\doibase 10.1103/PhysRevA.94.010102} {\bibfield  {journal} {\bibinfo
  {journal} {Phys. Rev. A}\ }\textbf {\bibinfo {volume} {94}},\ \bibinfo
  {pages} {010102} (\bibinfo {year} {2016})}\BibitemShut {NoStop}%
\bibitem [{\citenamefont {Steel}\ \emph {et~al.}(1998)\citenamefont {Steel},
  \citenamefont {Olsen}, \citenamefont {Plimak}, \citenamefont {Drummond},
  \citenamefont {Tan}, \citenamefont {Collett}, \citenamefont {Walls},\ and\
  \citenamefont {Graham}}]{Steel:1998}%
  \BibitemOpen
  \bibfield  {author} {\bibinfo {author} {\bibfnamefont {M.~J.}\ \bibnamefont
  {Steel}}, \bibinfo {author} {\bibfnamefont {M.~K.}\ \bibnamefont {Olsen}},
  \bibinfo {author} {\bibfnamefont {L.~I.}\ \bibnamefont {Plimak}}, \bibinfo
  {author} {\bibfnamefont {P.~D.}\ \bibnamefont {Drummond}}, \bibinfo {author}
  {\bibfnamefont {S.~M.}\ \bibnamefont {Tan}}, \bibinfo {author} {\bibfnamefont
  {M.~J.}\ \bibnamefont {Collett}}, \bibinfo {author} {\bibfnamefont {D.~F.}\
  \bibnamefont {Walls}}, \ and\ \bibinfo {author} {\bibfnamefont
  {R.}~\bibnamefont {Graham}},\ }\href {\doibase 10.1103/PhysRevA.58.4824}
  {\bibfield  {journal} {\bibinfo  {journal} {Phys. Rev. A}\ }\textbf {\bibinfo
  {volume} {58}},\ \bibinfo {pages} {4824} (\bibinfo {year}
  {1998})}\BibitemShut {NoStop}%
\bibitem [{\citenamefont {Blakie}\ \emph {et~al.}(2008)\citenamefont {Blakie},
  \citenamefont {Bradley}, \citenamefont {Davis}, \citenamefont {Ballagh},\
  and\ \citenamefont {Gardiner}}]{Blakie:2008}%
  \BibitemOpen
  \bibfield  {author} {\bibinfo {author} {\bibfnamefont {P.~B.}\ \bibnamefont
  {Blakie}}, \bibinfo {author} {\bibfnamefont {A.~S.}\ \bibnamefont {Bradley}},
  \bibinfo {author} {\bibfnamefont {M.~J.}\ \bibnamefont {Davis}}, \bibinfo
  {author} {\bibfnamefont {R.~J.}\ \bibnamefont {Ballagh}}, \ and\ \bibinfo
  {author} {\bibfnamefont {C.~W.}\ \bibnamefont {Gardiner}},\ }\bibfield
  {booktitle} {\emph {\bibinfo {booktitle} {Advances in Physics}},\ }\href
  {\doibase 10.1080/00018730802564254} {\bibfield  {journal} {\bibinfo
  {journal} {Advances in Physics}\ }\textbf {\bibinfo {volume} {57}},\ \bibinfo
  {pages} {363} (\bibinfo {year} {2008})}\BibitemShut {NoStop}%
\bibitem [{\citenamefont {Polkovnikov}(2010)}]{Polkovnikov:2010}%
  \BibitemOpen
  \bibfield  {author} {\bibinfo {author} {\bibfnamefont {A.}~\bibnamefont
  {Polkovnikov}},\ }\href {\doibase
  http://dx.doi.org/10.1016/j.aop.2010.02.006} {\bibfield  {journal} {\bibinfo
  {journal} {Annals of Physics}\ }\textbf {\bibinfo {volume} {325}},\ \bibinfo
  {pages} {1790} (\bibinfo {year} {2010})}\BibitemShut {NoStop}%
\bibitem [{\citenamefont {Opanchuk}\ and\ \citenamefont
  {Drummond}(2013)}]{Opanchuk:2013}%
  \BibitemOpen
  \bibfield  {author} {\bibinfo {author} {\bibfnamefont {B.}~\bibnamefont
  {Opanchuk}}\ and\ \bibinfo {author} {\bibfnamefont {P.~D.}\ \bibnamefont
  {Drummond}},\ }\href {\doibase 10.1063/1.4801781} {\bibfield  {journal}
  {\bibinfo  {journal} {Journal of Mathematical Physics}\ }\textbf {\bibinfo
  {volume} {54}},\ \bibinfo {eid} {042107} (\bibinfo {year}
  {2013})}\BibitemShut {NoStop}%
\bibitem [{\citenamefont {Walls}\ and\ \citenamefont
  {Milburn}(2008)}]{Walls:2008}%
  \BibitemOpen
  \bibfield  {author} {\bibinfo {author} {\bibfnamefont {D.~F.}\ \bibnamefont
  {Walls}}\ and\ \bibinfo {author} {\bibfnamefont {G.~J.}\ \bibnamefont
  {Milburn}},\ }\href@noop {} {\emph {\bibinfo {title} {Quantum Optics}}},\
  \bibinfo {edition} {2nd}\ ed.\ (\bibinfo  {publisher} {Springer-Verlag},\
  \bibinfo {address} {Berlin and Heidelberg},\ \bibinfo {year}
  {2008})\BibitemShut {NoStop}%
\bibitem [{\citenamefont {Gardiner}\ and\ \citenamefont
  {Zoller}(2004)}]{Gardiner:2004b}%
  \BibitemOpen
  \bibfield  {author} {\bibinfo {author} {\bibfnamefont {C.~W.}\ \bibnamefont
  {Gardiner}}\ and\ \bibinfo {author} {\bibfnamefont {P.}~\bibnamefont
  {Zoller}},\ }\href@noop {} {\emph {\bibinfo {title} {Quantum Noise: A
  Handbook of Markovian and Non-Markovian Quantum Stochastic Methods with
  Applications to Quantum Optics}}},\ \bibinfo {edition} {3rd}\ ed.\ (\bibinfo
  {publisher} {Springer},\ \bibinfo {address} {Berlin and Heidelberg},\
  \bibinfo {year} {2004})\BibitemShut {NoStop}%
\bibitem [{\citenamefont {Sinatra}\ \emph {et~al.}(2002)\citenamefont
  {Sinatra}, \citenamefont {Lobo},\ and\ \citenamefont
  {Castin}}]{Sinatra:2002}%
  \BibitemOpen
  \bibfield  {author} {\bibinfo {author} {\bibfnamefont {A.}~\bibnamefont
  {Sinatra}}, \bibinfo {author} {\bibfnamefont {C.}~\bibnamefont {Lobo}}, \
  and\ \bibinfo {author} {\bibfnamefont {Y.}~\bibnamefont {Castin}},\ }\href
  {http://stacks.iop.org/0953-4075/35/i=17/a=301} {\bibfield  {journal}
  {\bibinfo  {journal} {Journal of Physics B: Atomic, Molecular and Optical
  Physics}\ }\textbf {\bibinfo {volume} {35}},\ \bibinfo {pages} {3599}
  (\bibinfo {year} {2002})}\BibitemShut {NoStop}%
\bibitem [{\citenamefont {Opanchuk}\ \emph {et~al.}(2012)\citenamefont
  {Opanchuk}, \citenamefont {Egorov}, \citenamefont {Hoffmann}, \citenamefont
  {Sidorov},\ and\ \citenamefont {Drummond}}]{Opanchuk:2012}%
  \BibitemOpen
  \bibfield  {author} {\bibinfo {author} {\bibfnamefont {B.}~\bibnamefont
  {Opanchuk}}, \bibinfo {author} {\bibfnamefont {M.}~\bibnamefont {Egorov}},
  \bibinfo {author} {\bibfnamefont {S.}~\bibnamefont {Hoffmann}}, \bibinfo
  {author} {\bibfnamefont {A.~I.}\ \bibnamefont {Sidorov}}, \ and\ \bibinfo
  {author} {\bibfnamefont {P.~D.}\ \bibnamefont {Drummond}},\ }\href
  {http://stacks.iop.org/0295-5075/97/i=5/a=50003} {\bibfield  {journal}
  {\bibinfo  {journal} {EPL (Europhysics Letters)}\ }\textbf {\bibinfo {volume}
  {97}},\ \bibinfo {pages} {50003} (\bibinfo {year} {2012})}\BibitemShut
  {NoStop}%
\bibitem [{\citenamefont {Olsen}\ and\ \citenamefont
  {Bradley}(2009)}]{Olsen:2009}%
  \BibitemOpen
  \bibfield  {author} {\bibinfo {author} {\bibfnamefont {M.}~\bibnamefont
  {Olsen}}\ and\ \bibinfo {author} {\bibfnamefont {A.}~\bibnamefont
  {Bradley}},\ }\href {\doibase 10.1016/j.optcom.2009.06.033} {\bibfield
  {journal} {\bibinfo  {journal} {Optics Communications}\ }\textbf {\bibinfo
  {volume} {282}},\ \bibinfo {pages} {3924 } (\bibinfo {year}
  {2009})}\BibitemShut {NoStop}%
\bibitem [{\citenamefont {Drummond}\ and\ \citenamefont
  {Hardman}(1993)}]{Drummond:1993}%
  \BibitemOpen
  \bibfield  {author} {\bibinfo {author} {\bibfnamefont {P.~D.}\ \bibnamefont
  {Drummond}}\ and\ \bibinfo {author} {\bibfnamefont {A.~D.}\ \bibnamefont
  {Hardman}},\ }\href {http://stacks.iop.org/0295-5075/21/i=3/a=005} {\bibfield
   {journal} {\bibinfo  {journal} {EPL (Europhysics Letters)}\ }\textbf
  {\bibinfo {volume} {21}},\ \bibinfo {pages} {279} (\bibinfo {year}
  {1993})}\BibitemShut {NoStop}%
\bibitem [{\citenamefont {Carter}(1995)}]{Carter:1995}%
  \BibitemOpen
  \bibfield  {author} {\bibinfo {author} {\bibfnamefont {S.~J.}\ \bibnamefont
  {Carter}},\ }\href {\doibase 10.1103/PhysRevA.51.3274} {\bibfield  {journal}
  {\bibinfo  {journal} {Phys. Rev. A}\ }\textbf {\bibinfo {volume} {51}},\
  \bibinfo {pages} {3274} (\bibinfo {year} {1995})}\BibitemShut {NoStop}%
\bibitem [{\citenamefont {Olsen}\ and\ \citenamefont
  {Davis}(2006)}]{Olsen:2006}%
  \BibitemOpen
  \bibfield  {author} {\bibinfo {author} {\bibfnamefont {M.~K.}\ \bibnamefont
  {Olsen}}\ and\ \bibinfo {author} {\bibfnamefont {M.~J.}\ \bibnamefont
  {Davis}},\ }\href {\doibase 10.1103/PhysRevA.73.063618} {\bibfield  {journal}
  {\bibinfo  {journal} {Phys. Rev. A}\ }\textbf {\bibinfo {volume} {73}},\
  \bibinfo {pages} {063618} (\bibinfo {year} {2006})}\BibitemShut {NoStop}%
\bibitem [{Dep()}]{DepletionFootnote}%
  \BibitemOpen
  \href@noop {} {}\bibinfo {howpublished} {For high depletion
  $\text{Var}(\hat{N}_s) \sim \langle \hat{N}_s \rangle^2$ no longer holds,
  which reduces the QFI.}\BibitemShut {Stop}%
\bibitem [{\citenamefont {Nolan}\ \emph {et~al.}()\citenamefont {Nolan},
  \citenamefont {Szigeti},\ and\ \citenamefont {Haine}}]{Nolan_in_prep}%
  \BibitemOpen
  \bibfield  {author} {\bibinfo {author} {\bibfnamefont {S.~P.}\ \bibnamefont
  {Nolan}}, \bibinfo {author} {\bibfnamefont {S.~S.}\ \bibnamefont {Szigeti}},
  \ and\ \bibinfo {author} {\bibfnamefont {S.~A.}\ \bibnamefont {Haine}},\
  }\href@noop {} {}\bibinfo {howpublished} {In preparation}\BibitemShut
  {NoStop}%
\bibitem [{\citenamefont {Kasevich}\ and\ \citenamefont
  {Chu}(1992)}]{Kasevich:1992}%
  \BibitemOpen
  \bibfield  {author} {\bibinfo {author} {\bibfnamefont {M.}~\bibnamefont
  {Kasevich}}\ and\ \bibinfo {author} {\bibfnamefont {S.}~\bibnamefont {Chu}},\
  }\href {http://dx.doi.org/10.1007/BF00325375} {\bibfield  {journal} {\bibinfo
   {journal} {Applied Physics B: Lasers and Optics}\ }\textbf {\bibinfo
  {volume} {54}},\ \bibinfo {pages} {321} (\bibinfo {year} {1992})},\ \bibinfo
  {note} {10.1007/BF00325375}\BibitemShut {NoStop}%
\bibitem [{\citenamefont {Szigeti}\ \emph {et~al.}(2012)\citenamefont
  {Szigeti}, \citenamefont {Debs}, \citenamefont {Hope}, \citenamefont
  {Robins},\ and\ \citenamefont {Close}}]{Szigeti:2012}%
  \BibitemOpen
  \bibfield  {author} {\bibinfo {author} {\bibfnamefont {S.~S.}\ \bibnamefont
  {Szigeti}}, \bibinfo {author} {\bibfnamefont {J.~E.}\ \bibnamefont {Debs}},
  \bibinfo {author} {\bibfnamefont {J.~J.}\ \bibnamefont {Hope}}, \bibinfo
  {author} {\bibfnamefont {N.~P.}\ \bibnamefont {Robins}}, \ and\ \bibinfo
  {author} {\bibfnamefont {J.~D.}\ \bibnamefont {Close}},\ }\href
  {http://stacks.iop.org/1367-2630/14/i=2/a=023009} {\bibfield  {journal}
  {\bibinfo  {journal} {New Journal of Physics}\ }\textbf {\bibinfo {volume}
  {14}},\ \bibinfo {pages} {023009} (\bibinfo {year} {2012})}\BibitemShut
  {NoStop}%
\bibitem [{\citenamefont {Robins}\ \emph {et~al.}(2013)\citenamefont {Robins},
  \citenamefont {Altin}, \citenamefont {Debs},\ and\ \citenamefont
  {Close}}]{Robins:2013}%
  \BibitemOpen
  \bibfield  {author} {\bibinfo {author} {\bibfnamefont {N.~P.}\ \bibnamefont
  {Robins}}, \bibinfo {author} {\bibfnamefont {P.~A.}\ \bibnamefont {Altin}},
  \bibinfo {author} {\bibfnamefont {J.~E.}\ \bibnamefont {Debs}}, \ and\
  \bibinfo {author} {\bibfnamefont {J.~D.}\ \bibnamefont {Close}},\ }\bibfield
  {booktitle} {\emph {\bibinfo {booktitle} {Atom lasers: production, properties
  and prospects for precision inertial measurement}},\ }\href {\doibase
  http://dx.doi.org/10.1016/j.physrep.2013.03.006} {\bibfield  {journal}
  {\bibinfo  {journal} {Physics Reports}\ }\textbf {\bibinfo {volume} {529}},\
  \bibinfo {pages} {265} (\bibinfo {year} {2013})}\BibitemShut {NoStop}%
\bibitem [{\citenamefont {Szigeti}\ \emph {et~al.}(2014)\citenamefont
  {Szigeti}, \citenamefont {Tonekaboni}, \citenamefont {Lau}, \citenamefont
  {Hood},\ and\ \citenamefont {Haine}}]{Szigeti:2014b}%
  \BibitemOpen
  \bibfield  {author} {\bibinfo {author} {\bibfnamefont {S.~S.}\ \bibnamefont
  {Szigeti}}, \bibinfo {author} {\bibfnamefont {B.}~\bibnamefont {Tonekaboni}},
  \bibinfo {author} {\bibfnamefont {W.~Y.~S.}\ \bibnamefont {Lau}}, \bibinfo
  {author} {\bibfnamefont {S.~N.}\ \bibnamefont {Hood}}, \ and\ \bibinfo
  {author} {\bibfnamefont {S.~A.}\ \bibnamefont {Haine}},\ }\href {\doibase
  10.1103/PhysRevA.90.063630} {\bibfield  {journal} {\bibinfo  {journal} {Phys.
  Rev. A}\ }\textbf {\bibinfo {volume} {90}},\ \bibinfo {pages} {063630}
  (\bibinfo {year} {2014})}\BibitemShut {NoStop}%
\bibitem [{\citenamefont {Tonekaboni}\ \emph {et~al.}(2015)\citenamefont
  {Tonekaboni}, \citenamefont {Haine},\ and\ \citenamefont
  {Szigeti}}]{Tonekaboni:2015}%
  \BibitemOpen
  \bibfield  {author} {\bibinfo {author} {\bibfnamefont {B.}~\bibnamefont
  {Tonekaboni}}, \bibinfo {author} {\bibfnamefont {S.~A.}\ \bibnamefont
  {Haine}}, \ and\ \bibinfo {author} {\bibfnamefont {S.~S.}\ \bibnamefont
  {Szigeti}},\ }\href {\doibase 10.1103/PhysRevA.91.033616} {\bibfield
  {journal} {\bibinfo  {journal} {Phys. Rev. A}\ }\textbf {\bibinfo {volume}
  {91}},\ \bibinfo {pages} {033616} (\bibinfo {year} {2015})}\BibitemShut
  {NoStop}%
\bibitem [{\citenamefont {Haine}\ \emph {et~al.}(2015)\citenamefont {Haine},
  \citenamefont {Szigeti}, \citenamefont {Lang},\ and\ \citenamefont
  {Caves}}]{Haine:2015}%
  \BibitemOpen
  \bibfield  {author} {\bibinfo {author} {\bibfnamefont {S.~A.}\ \bibnamefont
  {Haine}}, \bibinfo {author} {\bibfnamefont {S.~S.}\ \bibnamefont {Szigeti}},
  \bibinfo {author} {\bibfnamefont {M.~D.}\ \bibnamefont {Lang}}, \ and\
  \bibinfo {author} {\bibfnamefont {C.~M.}\ \bibnamefont {Caves}},\ }\href
  {\doibase 10.1103/PhysRevA.91.041802} {\bibfield  {journal} {\bibinfo
  {journal} {Phys. Rev. A}\ }\textbf {\bibinfo {volume} {91}},\ \bibinfo
  {pages} {041802} (\bibinfo {year} {2015})}\BibitemShut {NoStop}%
\bibitem [{\citenamefont {Haine}\ and\ \citenamefont
  {Szigeti}(2015)}]{Haine:2015b}%
  \BibitemOpen
  \bibfield  {author} {\bibinfo {author} {\bibfnamefont {S.~A.}\ \bibnamefont
  {Haine}}\ and\ \bibinfo {author} {\bibfnamefont {S.~S.}\ \bibnamefont
  {Szigeti}},\ }\href {\doibase 10.1103/PhysRevA.92.032317} {\bibfield
  {journal} {\bibinfo  {journal} {Phys. Rev. A}\ }\textbf {\bibinfo {volume}
  {92}},\ \bibinfo {pages} {032317} (\bibinfo {year} {2015})}\BibitemShut
  {NoStop}%
\bibitem [{\citenamefont {Tojo}\ \emph {et~al.}(2009)\citenamefont {Tojo},
  \citenamefont {Hayashi}, \citenamefont {Tanabe}, \citenamefont {Hirano},
  \citenamefont {Kawaguchi}, \citenamefont {Saito},\ and\ \citenamefont
  {Ueda}}]{Tojo:2009}%
  \BibitemOpen
  \bibfield  {author} {\bibinfo {author} {\bibfnamefont {S.}~\bibnamefont
  {Tojo}}, \bibinfo {author} {\bibfnamefont {T.}~\bibnamefont {Hayashi}},
  \bibinfo {author} {\bibfnamefont {T.}~\bibnamefont {Tanabe}}, \bibinfo
  {author} {\bibfnamefont {T.}~\bibnamefont {Hirano}}, \bibinfo {author}
  {\bibfnamefont {Y.}~\bibnamefont {Kawaguchi}}, \bibinfo {author}
  {\bibfnamefont {H.}~\bibnamefont {Saito}}, \ and\ \bibinfo {author}
  {\bibfnamefont {M.}~\bibnamefont {Ueda}},\ }\href {\doibase
  10.1103/PhysRevA.80.042704} {\bibfield  {journal} {\bibinfo  {journal} {Phys.
  Rev. A}\ }\textbf {\bibinfo {volume} {80}},\ \bibinfo {pages} {042704}
  (\bibinfo {year} {2009})}\BibitemShut {NoStop}%
\bibitem [{\citenamefont {L{\"u}cke}\ \emph {et~al.}(2011)\citenamefont
  {L{\"u}cke}, \citenamefont {Scherer}, \citenamefont {Kruse}, \citenamefont
  {Pezz{\'e}}, \citenamefont {Deuretzbacher}, \citenamefont {Hyllus},
  \citenamefont {Topic}, \citenamefont {Peise}, \citenamefont {Ertmer},
  \citenamefont {Arlt}, \citenamefont {Santos}, \citenamefont {Smerzi},\ and\
  \citenamefont {Klempt}}]{Lucke:2011}%
  \BibitemOpen
  \bibfield  {author} {\bibinfo {author} {\bibfnamefont {B.}~\bibnamefont
  {L{\"u}cke}}, \bibinfo {author} {\bibfnamefont {M.}~\bibnamefont {Scherer}},
  \bibinfo {author} {\bibfnamefont {J.}~\bibnamefont {Kruse}}, \bibinfo
  {author} {\bibfnamefont {L.}~\bibnamefont {Pezz{\'e}}}, \bibinfo {author}
  {\bibfnamefont {F.}~\bibnamefont {Deuretzbacher}}, \bibinfo {author}
  {\bibfnamefont {P.}~\bibnamefont {Hyllus}}, \bibinfo {author} {\bibfnamefont
  {O.}~\bibnamefont {Topic}}, \bibinfo {author} {\bibfnamefont
  {J.}~\bibnamefont {Peise}}, \bibinfo {author} {\bibfnamefont
  {W.}~\bibnamefont {Ertmer}}, \bibinfo {author} {\bibfnamefont
  {J.}~\bibnamefont {Arlt}}, \bibinfo {author} {\bibfnamefont {L.}~\bibnamefont
  {Santos}}, \bibinfo {author} {\bibfnamefont {A.}~\bibnamefont {Smerzi}}, \
  and\ \bibinfo {author} {\bibfnamefont {C.}~\bibnamefont {Klempt}},\ }\href
  {http://www.sciencemag.org/content/334/6057/773.abstract N2 - Interferometers
  with atomic ensembles are an integral part of modern precision metrology.
  However, these interferometers are fundamentally restricted by the shot noise
  limit, which can only be overcome by creating quantum entanglement among the
  atoms. We used spin dynamics in Bose-Einstein condensates to create large
  ensembles of up to 104 pair-correlated atoms with an interferometric
  sensitivity decibels beyond the shot noise limit. Our proof-of-principle
  results point the way toward a new generation of atom interferometers.}
  {\bibfield  {journal} {\bibinfo  {journal} {Science}\ }\textbf {\bibinfo
  {volume} {334}},\ \bibinfo {pages} {773} (\bibinfo {year}
  {2011})}\BibitemShut {NoStop}%
\bibitem [{\citenamefont {Gross}\ \emph {et~al.}(2011)\citenamefont {Gross},
  \citenamefont {Strobel}, \citenamefont {Nicklas}, \citenamefont {Zibold},
  \citenamefont {Bar-Gill}, \citenamefont {Kurizki},\ and\ \citenamefont
  {Oberthaler}}]{Gross:2011}%
  \BibitemOpen
  \bibfield  {author} {\bibinfo {author} {\bibfnamefont {C.}~\bibnamefont
  {Gross}}, \bibinfo {author} {\bibfnamefont {H.}~\bibnamefont {Strobel}},
  \bibinfo {author} {\bibfnamefont {E.}~\bibnamefont {Nicklas}}, \bibinfo
  {author} {\bibfnamefont {T.}~\bibnamefont {Zibold}}, \bibinfo {author}
  {\bibfnamefont {N.}~\bibnamefont {Bar-Gill}}, \bibinfo {author}
  {\bibfnamefont {G.}~\bibnamefont {Kurizki}}, \ and\ \bibinfo {author}
  {\bibfnamefont {M.~K.}\ \bibnamefont {Oberthaler}},\ }\href
  {http://dx.doi.org/10.1038/nature10654} {\bibfield  {journal} {\bibinfo
  {journal} {Nature}\ }\textbf {\bibinfo {volume} {480}},\ \bibinfo {pages}
  {219} (\bibinfo {year} {2011})}\BibitemShut {NoStop}%
\bibitem [{\citenamefont {Davis}\ \emph {et~al.}(2016)\citenamefont {Davis},
  \citenamefont {Bentsen},\ and\ \citenamefont {Schleier-Smith}}]{Davis:2016}%
  \BibitemOpen
  \bibfield  {author} {\bibinfo {author} {\bibfnamefont {E.}~\bibnamefont
  {Davis}}, \bibinfo {author} {\bibfnamefont {G.}~\bibnamefont {Bentsen}}, \
  and\ \bibinfo {author} {\bibfnamefont {M.}~\bibnamefont {Schleier-Smith}},\
  }\href {\doibase 10.1103/PhysRevLett.116.053601} {\bibfield  {journal}
  {\bibinfo  {journal} {Phys. Rev. Lett.}\ }\textbf {\bibinfo {volume} {116}},\
  \bibinfo {pages} {053601} (\bibinfo {year} {2016})}\BibitemShut {NoStop}%
\bibitem [{\citenamefont {Dennis}\ \emph {et~al.}(2013)\citenamefont {Dennis},
  \citenamefont {Hope},\ and\ \citenamefont {Johnsson}}]{Dennis:2012}%
  \BibitemOpen
  \bibfield  {author} {\bibinfo {author} {\bibfnamefont {G.~R.}\ \bibnamefont
  {Dennis}}, \bibinfo {author} {\bibfnamefont {J.~J.}\ \bibnamefont {Hope}}, \
  and\ \bibinfo {author} {\bibfnamefont {M.~T.}\ \bibnamefont {Johnsson}},\
  }\href {\doibase 10.1016/j.cpc.2012.08.016} {\bibfield  {journal} {\bibinfo
  {journal} {Computer Physics Communications}\ }\textbf {\bibinfo {volume}
  {184}},\ \bibinfo {pages} {201} (\bibinfo {year} {2013})}\BibitemShut
  {NoStop}%
\end{thebibliography}%

\begin{widetext}

\renewcommand{\thefigure}{[S\arabic{figure}]} 
\renewcommand{\theequation}{S\arabic{equation}} 
\renewcommand\labelenumi{[\theenumi]}

\setcounter{equation}{0}
\setcounter{figure}{0}
\newpage

\section{Supplemental material: Pumped-up SU(1,1) interferometry}

In this supplemental material we provide further details on the calculation of key results reported in the main text, including the quantum Fisher information and phase sensitivity of a number-sum measurement (with and without losses), as well as a brief description of our truncated Wigner simulations.

\section{Pump-enhanced SU(1,1) interferometry with a spinor Bose-Einstein condensate}

\subsection{Quantum Fisher Information}
Consider a pure three-mode input state of the form
\begin{equation}
	|\psi_0 \rangle = \sum_{N=0}^\infty c_N |N,0,0 \rangle. \label{input_state}
\end{equation}
Here the Fock state $|n_0, n_+, n_- \rangle$ corresponds to the pump mode $\hat{a}_0$ with particle number $n_0$ and the side modes $\hat{a}_\pm$ with occupation numbers $n_\pm$. We assume that the nonlinear process that transfers correlated pairs of particles from the pump to the side modes can be described by the map:
\begin{equation}
	|N,0,0\rangle \mapsto \sum_{k \leq N/2} d_k(N) |N - 2k, k, k\rangle, \label{Fock_mappy}
\end{equation}
for some set of complex coefficients $d_k(N)$. The spin-mixing Hamiltonian $\hat{H}_\text{SMD} = \hbar \kappa ( \hat{a}_0^2 \hat{a}_+^\dag \hat{a}_-^\dag + h.c.)$ is a special case of this map. The initial state Eq.~(\ref{input_state}) is therefore mapped to
\begin{equation}
	|\psi_1 \rangle = \sum_{N=0}^\infty \sum_{n \leq N/2} C_n(N) |N-2n,n,n\rangle,
\end{equation}
where $C_n(N) \equiv c_N d_n(N)$. This state is then `pumped-up' by passing it through a tritter, described by the Hamiltonian
\begin{equation}
	\hat{H}_\text{tr} = \tfrac{\hbar \Omega}{\sqrt{2}}\big[ e^{i \vartheta}\hat{a}_0^\dag(\hat{a}_+ + \hat{a}_-) + e^{-i \vartheta} \hat{a}_0(\hat{a}_+^\dag + \hat{a}_-^\dag) \big]. \label{tritter_Ham}
\end{equation}
Since this is a linear process, the evolution can be analytically solved in the Heisenberg picture:
\begin{subequations}
\label{HP_tritter}
\begin{align}
	\hat{a}_\pm(\theta) 	&= \hat{a}_\pm \cos^2(\tfrac{\theta}{2}) - \hat{a}_\mp \sin^2(\tfrac{\theta}{2}) - \tfrac{i e^{-i \vartheta}}{\sqrt{2}} \hat{a}_0 \sin \theta, \\
	\hat{a}_0(\theta)	&= \hat{a}_0 \cos \theta - \tfrac{i e^{i \vartheta}}{\sqrt{2}} ( \hat{a}_+ + \hat{a}_- ) \sin \theta,
\end{align}
\end{subequations}
where $\theta = \Omega t$ for evolution time $t$. After this tritter, both side modes undergo a phase shift $\phi/2$, corresponding to the unitary $\hat{U}_{\phi} = \exp(-i \phi \hat{N}_s/2)$ where $\hat{N}_s = \hat{N}_+ + \hat{N}_-$ is the number sum operator. That is, $\hat{N}_s/2$ is the generator of the phase shift $\phi$, which is the classical parameter we wish to estimate. Since all subsequent operations can be conceptualized as part of the measurement process, they do not change the quantum Fisher information (QFI), which is given by the expression 
\begin{align}
	\mathcal{F} 	&= 4 \text{Var}(\hat{N}_s /2) = \langle \psi_1 | \hat{N}_s(\theta)^2 | \psi_1 \rangle - \langle \psi_1 | \hat{N}_s(\theta) | \psi_1 \rangle^2, \label{QFI_expression}
\end{align}
where $\hat{N}_s(\theta) = \hat{a}_+^\dag(\theta) \hat{a}_+(\theta) + \hat{a}_-^\dag(\theta) \hat{a}_-(\theta)$. The number-conserving process~(\ref{Fock_mappy}) implies that all non-number conserving expectations (e.g. $\langle \hat{a}_0^2 \hat{a}_- \rangle$) and all expectations that give different numbers of particles in the side modes (e.g. $\langle \hat{a}_0^\dag \hat{a}_+^\dag \hat{a}_-^2 \rangle$) are zero. This simplification yields
\begin{subequations}
\begin{align}
	\langle \hat{N}_s(\theta) \rangle 	&= \langle \hat{N}_0 \rangle \sin^2 \theta + \tfrac{1}{2} \langle \hat{N}_s \rangle \left( 1 + \cos^2 \theta \right), \\
	\langle \hat{N}_s(\theta)^2 \rangle 	&= \left[ \langle \hat{N}_0 \rangle - A(\vartheta)\right] \cos^2 \theta \sin^2 \theta + \langle \hat{N}_0^2\rangle \sin^4 \theta +  \langle \hat{N}_0 \hat{N}_s \rangle \left( 1 + 2\cos^2\theta \right) \sin^2 \theta \notag \\
								&+ \tfrac{1}{4} \langle \hat{N}_s \rangle \left( 1 + \cos^2\theta \right) \sin^2 \theta + \tfrac{1}{32} \langle \hat{N}_s^2 \rangle \left[ 20 \cos^2 \theta + 3 \left( 3 + \cos^2(2 \theta) \right) \right] .
\end{align}
\end{subequations}
where $\hat{N}_0 = \hat{a}_0^\dag \hat{a}_0$, $A(\vartheta) = \exp(-2 i \vartheta)\langle \hat{a}_0^2\hat{a}_+^\dag \hat{a}_-^\dag \rangle + h.c.$, and the above expectations are all taken with respect to $|\psi_1\rangle$. The QFI is therefore
\begin{align}
	\mathcal{F}(\theta)	&= \tfrac{1}{4}(1+\cos^2\theta)^2 \text{Var}(\hat{N}_s) + \big[ \text{Var}(\hat{N}_0) + \tfrac{1}{4}\left(\tfrac{1}{2}\langle \hat{N}_s^2 \rangle + \langle \hat{N}_s\rangle\right)\big] \sin^4 \theta \notag \\
						&+ (1 + 2 \cos^2 \theta) \sin^2 \theta\,  \text{Cov}(\hat{N}_0, \hat{N}_s) + \tfrac{1}{4}\big[\langle \hat{N}_0\rangle(\langle \hat{N}_s\rangle+1) + \tfrac{1}{2}\langle \hat{N}_s \rangle - A(\vartheta)\big] \sin^2(2\theta).
	 \label{QFI_pure}
\end{align}

\subsubsection{Optimal tritter parameters}
We now determine the optimal tritter angle $\theta$ and phase $\vartheta$ that maximize the QFI. Let us begin with the latter. Since
\begin{align}
	\frac{\partial \mathcal{F}}{\partial \vartheta}	&=  \frac{i}{2}\left( e^{-2i \vartheta}\langle \hat{a}_0^2\hat{a}_+^\dag \hat{a}_-^\dag \rangle - e^{2i \vartheta}\langle (\hat{a}_0^\dag)^2\hat{a}_+ \hat{a}_- \rangle\right) \sin^2(2\theta),
\end{align}
this implies that two critical points occur (modulo $\pi$) at 
\begin{equation}
	\vartheta_\pm = \frac{i}{2} \ln \left( \pm \sqrt{\frac{\langle (\hat{a}_0^\dag)^2\hat{a}_+ \hat{a}_0 \rangle}{\langle \hat{a}_0^2\hat{a}_+^\dag \hat{a}_0^\dag \rangle}}\right).  
\end{equation}
The tritter phase only affects the QFI via $A(\vartheta)$; at the optimal phase $A(\vartheta_\pm) = \pm \,2 | \langle \hat{a}_0^2\hat{a}_+^\dag \hat{a}_0^\dag \rangle |$. Since $| \langle \hat{a}_0^2\hat{a}_+^\dag \hat{a}_0^\dag \rangle | \geq 0$, inspection of the QFI Eq.~(\ref{QFI_pure}) reveals that $\vartheta_-$ maximizes the QFI. 

We can similarly determine the optimal tritter angle:
\begin{align}
	\frac{\partial \mathcal{F}}{\partial \theta} 	&= \Big[ \tfrac{1}{8}\langle \hat{N}_s \rangle (\langle \hat{N}_s \rangle+2) - \tfrac{5}{8} \text{Var}(\hat{N}_s) + \text{Var}(\hat{N}_0) + \text{Cov}(\hat{N}_0 \hat{N}_s) + \Big( \langle \hat{N}_0 \rangle (\langle \hat{N}_s \rangle + 1)  - \text{Var}(\hat{N}_0) \notag \\
					&+ 2\text{Cov}(\hat{N}_0, \hat{N}_s) - \tfrac{3}{8}\text{Var}(\hat{N}_s) - \tfrac{1}{8}\langle \hat{N}_s \rangle(\langle \hat{N}_s \rangle-2) -  A(\vartheta_-)\Big)\cos(2\theta) \Big] \sin(2\theta).
\end{align}
Therefore, on the interval $\theta \in [0, \pi/2]$ the critical points are $\theta_\text{opt} = 0, \pi/2$ and $\theta_c^{(x_c)} = \cos^{-1}(x_c)/2$ for
\begin{equation}
	x_c = \frac{\langle \hat{N}_s\rangle(\langle \hat{N}_s\rangle+2)- 5 \text{Var}(\hat{N}_s) + 8 \text{Var}(\hat{N}_0) + 8 \text{Cov}(\hat{N}_0, \hat{N}_s)}{8 \left( \text{Var}(\hat{N}_0) - \langle \hat{N}_0 \rangle (\langle \hat{N}_s\rangle + 1) - 2 \text{Cov}(\hat{N}_0, \hat{N}_s) + A(\vartheta_-)\right) + \langle \hat{N}_s \rangle(\langle \hat{N}_s \rangle-2) + 3 \text{Var}(\hat{N}_s)}.
\end{equation}
This final critical point only exists if $|x_c| < 1$. Thus,
\begin{subequations}
\begin{align}
	\mathcal{F}(0) 		&=\text{Var}(\hat{N}_s),\\
	\mathcal{F}(\pi/2)		&= \text{Var}(\hat{N}_0) + \tfrac{3}{8}\text{Var}(\hat{N}_s) + \text{Cov}(\hat{N}_0, \hat{N}_s) + \tfrac{1}{8}\langle \hat{N}_s \rangle \big(\langle \hat{N}_s\rangle + 2\big), \\
	\mathcal{F}(\theta_c^{(x_c)})	&= \tfrac{1}{4}\big[ \text{Var}(\hat{N}_0) + \tfrac{19}{8}\text{Var}(\hat{N}_s) + 4 \text{Cov}(\hat{N}_0,\hat{N}_s) + \langle \hat{N}_s \rangle \big( \langle \hat{N}_0 \rangle + \tfrac{1}{8} (\langle \hat{N}_s \rangle + 2)\big) - A(\vartheta_-) \big] \notag \\
	& + \tfrac{x_c}{16}\big[ 8 \langle \hat{N}_0 \rangle + 5 \text{Var}(\hat{N}_s) - 8\text{Var}(\hat{N}_0) - 8\text{Cov}(\hat{N}_0, \hat{N}_s) - \langle \hat{N}_s \rangle\big(\langle \hat{N}_s \rangle+2\big)\big] \notag \\
	& + \tfrac{x_c^2}{4}\big[ \text{Var}(\hat{N}_0) + \tfrac{3}{8}\text{Var}(\hat{N}_s) - 2\text{Cov}(\hat{N}_0, \hat{N}_s) - \big(\langle \hat{N}_0 \rangle-\tfrac{1}{8}\langle \hat{N}_s \rangle\big)\big(\langle \hat{N}_s \rangle+2\big) + A(\vartheta_-) \big] \notag \\
	& -\tfrac{1}{8}(x_c - 1) \big[2 \langle \hat{N}_0 \rangle + \langle \hat{N}_s \rangle - \big(2\langle \hat{N}_0 \rangle - \langle \hat{N}_s \rangle\big)x_c\big],
\end{align}
\end{subequations}
and 
\begin{equation}
	\mathcal{F}(\theta_\text{opt}) = \begin{cases}
							\max \left\{ \mathcal{F}(0), \mathcal{F}(\pi/2), \mathcal{F}(\theta_c^{(x_c)})	\right\}, & |x_c| < 1 \\
							\max \left\{ \mathcal{F}(0), \mathcal{F}(\pi/2) \right\}, & |x_c| \geq 1
							\end{cases}
\end{equation}

\subsubsection{Undepleted pump regime}
We now assume that the pump $\hat{a}_0$ is initially in a coherent state with mean number $\overline{N}$ and phase $\vartheta_p$. Provided the average number of particles outcoupled from the pump to the side modes remains small compared with $\overline{N}$, we can assume that the pump remains in a coherent state with a largely unchanged number of particles (i.e. $\overline{N} \mapsto \overline{N} - \langle \hat{N}_s \rangle  \approx \overline{N}$). This is the undepleted pump approximation. Formally, we assume that all relevant expectations with respect to $|\psi_1\rangle$ are given by
\begin{subequations}
\begin{align}
	\langle \hat{N}_0 \rangle		&= \overline{N} - \mathcal{N}_s, \\
	\langle \hat{N}_0^2 \rangle	&= (\overline{N} - \mathcal{N}_s)(\overline{N} - \mathcal{N}_s - 1), \\
	\langle \hat{N}_s^2 \rangle	&= 2 \mathcal{N}_s (\mathcal{N}_s + 1), \\
	\langle \hat{N}_0 \hat{N}_s \rangle	&= (\overline{N} - \mathcal{N}_s) \mathcal{N}_s, \\
	 \langle \hat{a}_0^2\hat{a}_+^\dag \hat{a}_-^\dag \rangle	&= \frac{i}{2}e^{i(2 \vartheta_p + \vartheta_\text{sq})}(\overline{N} - \mathcal{N}_s) \sqrt{\mathcal{N}_s(\mathcal{N}_s+2)},
\end{align}
\end{subequations}
where $\mathcal{N}_s \equiv \langle \hat{N}_s \rangle$. This is consistent with an outcoupling process {$\hat{U}_\text{PA} = \exp\left[-i r\left(e^{i \vartheta_\text{sq}} \hat{a}_+ \hat{a}_- + h.c. \right)\right]$}, and therefore $\mathcal{N}_s = 2 \sinh^2 r$. Note that we have imposed the constraint $\overline{N} = \langle \hat{N}_0 \rangle + \langle \hat{N}_s \rangle$. The QFI Eq.~(\ref{QFI_pure}) therefore reduces to
\begin{align}
	\mathcal{F}(\theta)	&= \overline{N} \left[ 1 + \left( \mathcal{N}_s - \sqrt{\mathcal{N}_s(\mathcal{N}_s+2)} \sin  \nu \right) \cos^2 \theta \right] \sin^2 \theta \notag \\
	& + \frac{\mathcal{N}_s}{2}\left[ \mathcal{N}_s + \left( \mathcal{N}_s + 4 \right) \cos^2 \theta + \frac{1}{4}\left( 2\sqrt{\mathcal{N}_s(\mathcal{N}_s+2)}\sin \nu-3\mathcal{N}_s - 1 \right)\sin^2(2\theta) \right],
\end{align}
where $\nu = 2 (\vartheta - \vartheta_p) - \vartheta_\text{sq}$. The maximum QFI occurs for $\nu_\text{opt} = 3 \pi / 2$ and for tritter angle $\theta_\text{opt} = 0, \pi/2$ or $\theta_c^{(x_c)} = \cos^{-1}(x_c)/2$, where
\begin{equation}
	x_c = \frac{\mathcal{N}_s(\mathcal{N}_s+4)-2\overline{N}}{\mathcal{N}_s(2\overline{N} - 3 \mathcal{N}_s - 1)-2(\overline{N}-\mathcal{N}_s)\sqrt{\mathcal{N}_s(\mathcal{N}_s+2)} \sin \nu}.
\end{equation}
Assuming $\theta_c^{(x_c)}$ exists, $\nu = 3 \pi /2$, and $\overline{N} \gg 1$, an asymptotic expansion of $\theta_c^{(x_c)}$ in powers of $1/\overline{N}$ yields
\begin{equation}
	\theta_c^{(x_c)} = \frac{1}{4} \left( \pi + 2 \text{csc}^{-1}\left( \mathcal{N}_s + \sqrt{\mathcal{N}_s(\mathcal{N}_s+2)}\right) \right) + \mathcal{O}\left( 1 / \overline{N} \right). \label{approx_theta_c}
\end{equation}
Since $\text{csc}^{-1}(x)$ is undefined for $|x| < 1$, Eq.~(\ref{approx_theta_c}) implies that the critical point $\theta_c^{(x_c)}$ only exists for $\mathcal{N}_s \geq 1/4$. The QFI at the three critical points is
\begin{subequations}
\label{QFI_un_approx}
\begin{align}
	\mathcal{F}(0)		&= \mathcal{N}_s (\mathcal{N}_s + 2), \\
	\mathcal{F}(\pi/2)	&= \overline{N} + \tfrac{1}{2}\mathcal{N}_s^2, \\
	\mathcal{F}(\theta_c^{(x_c)})	&= \frac{ \overline{N}}{2}(1-x_c)\left[ 1 + \tfrac{1}{2}(1+x_c)\left(\mathcal{N}_s + \sqrt{\mathcal{N}_s(\mathcal{N}_s+2)}\right)\right] \notag \\
					&+ \frac{\mathcal{N}_s}{16} \left[ 15 + 9 \mathcal{N}_s + 4(\mathcal{N}_s+4)x_c + (3\mathcal{N}_s+1)(2x_c^2-1) + 4\sqrt{\mathcal{N}_s(\mathcal{N}_s+2)}(x_c^2-1)\right], \notag \\
					&= \frac{e^{2r}}{8}(1 + \coth r) \overline{N} + \mathcal{O}\left(\overline{N}^0\right). 
\end{align}
\end{subequations}
Eqs.~(\ref{QFI_un_approx}) are reported as Eq.~(7) in the main text. 



\subsection{Phase sensitivity for Loschmidt echo protocol restricted to side-mode measurements}

First we briefly review the Loschmidt echo protocol outlined in [32]; our pumped-up interferometer [see Fig.~1(b) of main text] evolves the initial state $|\psi_0\rangle$ to $|\psi_\text{out}\rangle = \hat{U}^\dag \exp(-i \phi \hat{N}_s/2) \hat{U} |\psi_0\rangle$, where $\hat{U} = \hat{U}_\text{tr}(\theta)\hat{U}_\text{SMD}$ first evolves the state via the nonlinear process Eq.~(\ref{Fock_mappy}) (i.e. spin-exchange collisions) and then according to the tritter operation Eqs.~(\ref{HP_tritter}). The measurement signal $\hat{\mathcal{S}}_\text{LE} = |\psi_0 \rangle \langle \psi_0|$ yields $P_0 \equiv \langle \psi_\text{out}| \hat{\mathcal{S}}_\text{LE} | \psi_\text{out} \rangle = 1 - \phi^2\mathcal{F}/4 + \mathcal{O}(\phi^4)$. Furthermore, the variance and slope are $\text{Var}(\hat{\mathcal{S}}_\text{LE}) = P_0 (1 - P_0) = \mathcal{F} \phi^2 / 4 + \mathcal{O}(\phi^4)$ and $(\partial_\phi P_0)^2 = \mathcal{F}^2 \phi^2 / 4 + \mathcal{O}(\phi^4)$, implying that when $\phi \to 0$ the sensitivity $(\Delta \phi)^2 = \text{Var}(\hat{\mathcal{S}}_\text{LE}) / (\partial_\phi \langle \hat{\mathcal{S}}_\text{LE} \rangle)^2 = 1 / \mathcal{F}$, thereby saturating the QCRB. 

Now suppose we restrict measurements to the side modes at the output, and we choose to project the output onto the initial (vacuum) side modes with measurement signal $\hat{\mathcal{S}}_\text{LE}' = |0,0 \rangle \langle 0,0| = \sum_N |N, 0,0 \rangle \langle N, 0,0|$. Then
\begin{align}
	P_0'	&\equiv \langle \psi_\text{out}| \hat{\mathcal{S}}_\text{LE}' | \psi_\text{out} \rangle \notag \\
		&=  \sum_{N,n,n'} d_n^*(N) d_{n'}(N) \langle \psi_1 | e^{+i \phi \hat{N}_s(\theta)/2} |N-2n',n',n' \rangle \langle N-2n, n, n | e^{-i \phi \hat{N}_s(\theta)/2} |\psi_1 \rangle \notag \\
		&= \sum_{N, M, M'} \sum_{n,n',m,m'} C_m^*(M) C_{m'}(M') d_n^*(N)d_{n'}^*(N) \notag \\
		&\times \langle M-2m,m,m| \left( 1 + \tfrac{i \phi}{2} \hat{N}_s(\theta) - \tfrac{\phi}{8} \hat{N}_s^2(\theta) + \mathcal{O}(\phi^3) \right) |N - 2n',n',n'\rangle \notag \\
		&\times \langle N-2n,n,n| \left( 1 - \tfrac{i \phi}{2} \hat{N}_s(\theta) - \tfrac{\phi}{8} \hat{N}_s^2(\theta)  + \mathcal{O}(\phi^3) \right) |M' - 2m',m',m'\rangle \notag \\
		&= 1 -  \frac{\phi^2}{4} \sum_N |c_N|^2 \left[ \sum_{n} |d_{n}(N)|^2 \mathcal{R}(N,n,\theta) - \left(\sum_{n'} |d_{n'}(N)|^2 \mathcal{Q}(N,n',\theta)\right)^2\right] + \mathcal{O}(\phi^3), \label{P0_dash}
\end{align}
where $\mathcal{Q}(N,n,\theta)$ and $\mathcal{R}(N,n,\theta)$ are defined via
\begin{subequations}
\begin{align}
	\langle N-2n,n,n | \hat{N}_s(\theta) | M - 2m, m ,m \rangle 	&= \left[N \sin^2 \theta + n \left( 1 + \cos^2 \theta \right)\right] \delta_{M,N} \delta_{n,m}, \notag\\
												&\equiv \mathcal{Q}(N,n,\theta) \delta_{M,N} \delta_{n,m},  \\
	\langle N-2n,n,n | \hat{N}_s(\theta)^2 | M - 2m, m ,m \rangle 	&= \Big\{ \left( N - \tilde{\mathcal{A}} \right) \cos^2 \theta \sin^2 \theta + N^2 \sin^4 \theta  \notag \\
								&+  2 N n \left( 1 + 2\cos^2\theta \right) \sin^2 \theta + \tfrac{1}{2} n \left( 1 + \cos^2\theta \right) \sin^2 \theta \notag \\
								&+ \tfrac{1}{8} n^2 \left[ 20 \cos^2 \theta + 3 \left( 3 + \cos^2(2 \theta) \right) \right] \Big\} \delta_{M,N} \delta_{n,m}, \notag\\
								&\equiv \mathcal{R}(N,n,\theta) \delta_{M,N} \delta_{n,m}, \label{Nsq_thing}
\end{align}
\end{subequations}
with $\tilde{\mathcal{A}} = (\langle N-2n,n,n | \hat{a}_0^2 \hat{a}_+^\dag \hat{a}_-^\dag | N - 2n, n ,n \rangle \exp(-2i \vartheta) + h.c.)$. The first term within the brackets of Eq.~(\ref{P0_dash}) gives
\begin{equation}
	\sum_N |c_N|^2 \sum_n |d_n(N)|^2 \mathcal{R}(N,n,\theta) = \langle \psi_1 | \hat{N}_s(\theta)^2 | \psi_1 \rangle,
\end{equation}
whereas the second yields
\begin{equation}
	\sum_N |c_N|^2 \left(\sum_{n} |d_{n}(N)|^2 \mathcal{Q}(N,n',\theta)\right)^2 = \langle \psi_1 | \hat{N}_s(\theta) | \psi_1 \rangle^2 + \left( \langle \psi_0 | \hat{N}_0^2 | \psi_0 \rangle - \langle \psi_0 |\hat{N}_0 | \psi_0 \rangle^2 \right) \sin^4 \theta.
\end{equation}
Consequently,
\begin{align}
	P_0'	&= 1 - \frac{\phi^2}{4}\left( \text{Var}(\hat{N}_s(\theta))|_{\psi_1} - \text{Var}(\hat{N}_0)|_{\psi_0} \sin^4 \theta \right) + \mathcal{O}(\phi^3).
\end{align}
where $\text{Var}(\hat{N}_0)|_{\psi_0} \equiv \langle \psi_0 | \hat{N}_0^2 | \psi_0 \rangle - \langle \psi_0 |\hat{N}_0 | \psi_0 \rangle^2$.
It follows that
\begin{align}
	\text{Var}(\hat{\mathcal{S}}_\text{LE}')	&= P_0' (1 - P_0') =\frac{\phi^2}{4}\left( \text{Var}(\hat{N}_s(\theta))|_{\psi_1} - \text{Var}(\hat{N}_0)|_{\psi_0} \sin^4 \theta \right) + \mathcal{O}(\phi^4) \\
	(\partial_\phi P_0')^2	&= \frac{\phi^2}{4}\left( \text{Var}(\hat{N}_s(\theta))|_{\psi_1} - \text{Var}(\hat{N}_0)|_{\psi_0} \sin^4 \theta \right)^2 + \mathcal{O}(\phi^4),
\end{align}
and therefore in the limit $\phi \to 0$
\begin{align}
	\Delta \phi	&= \frac{1}{\sqrt{\mathcal{F}(\theta) - \text{Var}(\hat{N}_0)|_{\psi_0} \sin^4 \theta}}.
\end{align}
We can see immediately that this will only saturate the QCRB if $\theta = 0$ and/or $\text{Var}(\hat{N}_0)|_{\psi_0} = 0$. However, for a coherent pump $\text{Var}(\hat{N}_0)|_{\psi_0} = \overline{N}$, leading to a sensitivity worse than the QCRB. More concretely, within the undepleted pump regime it is straightforward to show that the optimum tritter phase and angle is $\vartheta_\text{opt} = 3 \pi /2$ and $\theta_\text{opt} = \pi/4 + \mathcal{O}(1/\overline{N})$, respectively, yielding a minimum sensitivity of $\Delta \phi = 2 \exp(-r)/\overline{N} + \mathcal{O}(1/\overline{N}^{3/2})$. As shown below, this is identical to the minimum sensitivity obtained with the operationally simpler number-sum measurement.

\subsection{Phase sensitivity for number-sum measurement in undepleted pump regime}
Here we explicitly derive the phase sensitivity for the measurement signal $\hat{\mathcal{S}} = \hat{N}_s$ within the undepleted pump regime. This is most easily done by evolving the operators $\hat{a}_\pm$ in the Heisenberg picture. Specifically, we take our initial pump state to be a coherent state $| \alpha_0 e^{i \vartheta_p}\rangle$ where $\alpha_0 = \sqrt{\overline{N}}$ and $\hat{a}_\pm$ to initially be in vacuum;  all expectations will be taken with respect to this initial state $|\psi_0\rangle = |\alpha_0 e^{i \vartheta_p}, 0, 0\rangle$. The modes then undergo the follow stages of evolution:
\begin{enumerate}
	\item Parametric amplification, described by the unitary {$\hat{U}_\text{PA} = \exp\left[-i r\left(e^{i \vartheta_\text{sq}} \hat{a}_+ \hat{a}_- + h.c. \right)\right]$}:
	\begin{align}
		\hat{a}_\pm^{(1)}(r)	&\equiv \hat{U}_\text{PA}^\dag(r) \hat{a}_\pm \hat{U}_\text{PA}(r) = \hat{a}_\pm \cosh r - i e^{i \vartheta_\text{sq}} \hat{a}_\mp^\dag \sinh r.
	\end{align} 
	Under the undepleted pump approximation, we assume that $\hat{a}_0$ remains in a coherent state. However, we impose number conservation $\overline{N} = \langle \hat{N}_0 \rangle + \langle \hat{N}_+(r) \rangle + \langle \hat{N}_-(r) \rangle$. That is, we assume the pump coherent state amplitude evolves to $|\alpha(r)|^2 = |\alpha_0|^2 - 2 \sinh^2 r$.
	\item {First tritter described by Hamiltonian~(\ref{tritter_Ham}):
	\begin{subequations}
\begin{align}
	\hat{a}_\pm^{(2)}(r, \theta) 	&= \hat{a}_\pm^{(1)}(r) \cos^2(\tfrac{\theta}{2}) - \hat{a}_\mp^{(1)}(r) \sin^2(\tfrac{\theta}{2}) - \tfrac{ie^{-i \vartheta}}{\sqrt{2}} \hat{a}_0 \sin \theta, \\
	\hat{a}_0^{(2)}(r, \theta)	&= \hat{a}_0 \cos \theta - \tfrac{i e^{i \vartheta}}{\sqrt{2}}  \big( \hat{a}_+^{(1)}(r) + \hat{a}_-^{(1)}(r) \big) \sin \theta.
\end{align}
\end{subequations}}
	\item The unitary $\hat{U}(\phi) = \exp(-i \phi \hat{N}_s / 2)$ shifts the phase of the two side modes by $\phi/2$ relative to the pump mode:
\begin{equation}
	\hat{a}_\pm^{(3)}(r, \theta, \phi) = \hat{a}_\pm^{(2)}(r, \theta)  e^{-i \phi / 2}.
\end{equation}
	\item The second tritter with angle $-\theta$ (i.e. a $\pi$ phase shift relative to the first tritter):
	\begin{subequations}
	\begin{align}
	\hat{a}_\pm^{(4)}(r, \theta, \phi) 	&= \hat{a}_\pm^{(3)}(r, \theta, \phi) \cos^2(\tfrac{\theta}{2}) - \hat{a}_\mp^{(3)}(r, \theta, \phi) \sin^2(\tfrac{\theta}{2}) + \tfrac{ie^{-i \vartheta}}{\sqrt{2}} \hat{a}_0^{(2)}(r, \theta) \sin \theta, \\
	\hat{a}_0^{(4)}(r, \theta, \phi)	&=\hat{a}_0^{(2)}(r, \theta) \cos \theta + \tfrac{i e^{i \vartheta}}{\sqrt{2}} \big( \hat{a}_+^{(3)}(r, \theta, \phi) + \hat{a}_-^{(3)}(r, \theta, \phi) \big)\sin \theta.
\end{align}
\end{subequations}
	\item Finally, a second parametric amplification $\hat{U}_\text{PA}(-r)$ that reverses the evolution of the first parametric amplifier:
	\begin{align}
		\hat{a}_\pm^{(5)}(\theta, r, \phi)	&= \hat{a}_\pm^{(4)}(r, \theta, \phi) \cosh r + i e^{i \vartheta_\text{sq}} \big[\hat{a}_\mp^{(4)}(r, \theta, \phi)\big]^\dag \sinh r.
	\end{align} 
\end{enumerate}
We take $\hat{\mathcal{S}} = \big[\hat{a}_+^{(5)}(\theta, r, \phi)\big]^\dag \hat{a}_+^{(5)}(\theta, r, \phi) + \big[\hat{a}_-^{(5)}(\theta, r, \phi)\big]^\dag \hat{a}_-^{(5)}(\theta, r, \phi)$ as our measurement signal. By expressing $\hat{\mathcal{S}}$ in terms of $\hat{a}_\pm$ and $\hat{a}_0$, and taking expectations with respect to the initial state $|\psi_0\rangle$, we can show that
\begin{align}
	\langle \hat{\mathcal{S}} \rangle &= |\alpha(r)|^2 \sin^2(2\theta) \sin^2(\phi/4)\left[ \cosh(2r) - \sin(\nu + \phi/2) \sinh(2r)\right] \notag \\
							&+ \sin^2(\phi/4) \left[ \sin^2(2\theta)  4 \left( 3 + 4 \cos^2 \theta + \cos^2(2\theta)\right) \cos^2(\phi/4) \cosh^2 r \right] \sinh^2 r, \\
		\partial_{\phi}\langle \hat{\mathcal{S}} \rangle	&= \tfrac{1}{2}|\alpha(r)|^2 \sin^2(2\theta) \sin(\phi/4)\left[ \cos(\phi/4) \cosh(2r) - \sin (\nu + 3 \phi/4) \sinh(2r)\right] \notag \\
							&+ \cos(\phi/4) \left[ 2 \left( 3 + 4 \cos^2 \theta + \cos^2(2\theta) \right) \cos(\phi/2) \cosh^2 r + \tfrac{1}{2} \sin^2(2\theta)\right] \sin(\phi/4) \sinh^2 r, \\
		\langle \hat{\mathcal{S}}^2 \rangle	&= \frac{\phi^2}{16}\Big\{ |\alpha(r)|^2 \sin^2(2\theta) \left[ \cosh(2r) - \sin \nu \sinh(2r) \right] + \sin^2(2\theta) \sinh^2 r \notag \\
				&+ 2\left( 3 + 4 \cos^2 \theta + \cos^2(2\theta)\right) \sinh^2(2r) \Big\}+ \mathcal{O}(\phi^4).
\end{align}
We can determine \emph{a posteriori} that the optimal phase sensitivity occurs at $\phi = 0$, and
\begin{align}
	\Delta \phi_N	&= \frac{\sqrt{\text{Var}(\hat{\mathcal{S}})}}{|\partial_{\phi}\langle \hat{\mathcal{S}} \rangle|} \notag \\
			&= \frac{2 \sqrt{\left(|\alpha(r)|^2 \eta(r) + \sinh^2 r\right)\sin^2(2\theta) + 2 \left[ 3 + 4 \cos^2 \theta + \cos^2 (2\theta)\right] \sinh^2(2r)}}{\left| \left(|\alpha(r)|^2 \eta(r) + \sinh^2 r\right)\sin^2(2\theta) + \left[ 3 + 4 \cos^2 \theta + \cos^2 (2\theta)\right] \sinh^2(2r) \right|} \notag \\
					&= 2|\text{csc}(2\theta)|\left(\frac{1}{\sqrt{ \eta(r) \overline{N}}} + \frac{ \left( \eta(r) - 1/2\right)\sinh^2 r}{( \eta(r) \overline{N})^{3/2}}\right) + \mathcal{O}(1 / \overline{N}^{5/2}),
\end{align}
where $\eta(r) \equiv \cosh(2r) - \sin \nu \sinh(2r)$. This reveals that to leading order in $\overline{N}$ the optimal parameter choice is $\nu = 3 \pi /2$ and $\theta = \pi / 4$. Then the minimum sensitivity is $\Delta \phi_N \approx  2 \exp(-r) / \sqrt{\overline{N}}$. 

\subsubsection{Insensitivity to detection noise}

We model imperfect detection resolution as a Gaussian noise of variance $(\Delta n)^2$, which corresponds to an uncertainty $\Delta n$ in the number of atoms measured at the output. This technical noise adds (in quadrature) with the quantum noise on the signal. Consequently, assuming $\nu = 3 \pi /2$ and $\theta = \pi/4$:
\begin{align}
	\Delta \phi 	&= \frac{\sqrt{\text{Var}(\hat{\mathcal{S}}) + (\Delta n)^2}}{|\partial_{\phi}\langle\hat{\mathcal{S}} \rangle|} \notag \\
				&= 2\sqrt{\frac{\sec^2\left(\frac{\phi}{4}\right)+2 e^{3r} \sinh r \left[ 1 + 2 \cosh(2r) + \sin^2\left(\frac{\phi}{2}\right) \left(1 + \coth r \right) \sinh^2(2r) + 2 \sin^2\left(\frac{\phi}{4}\right) \sinh(4r)\right]}{\left[ 1 + \left( e^{4r} - 1 \right) \cos(\phi/2) \right]^2 \overline{N}}} \notag \\
				& + \mathcal{O}(1/\overline{N}^{3/2}).
\end{align}
We have kept $\phi$ general, as the inclusion of this noise could potentially shift the optimum operating point. However, for our pump-enhanced SU(1,1) interferometer, to leading order the optimal operating point remains $\phi = 0$, and the sensitivity is \emph{independent} of the number resolution - i.e. $\Delta \phi \approx  2 \exp(-r) / \sqrt{\overline{N}}$. This result is only true provided $\Delta n \lesssim \overline{N}$. 

In contrast, conventional SU(1,1) interferometry has a weak dependence on detection noise. Specifically,
\begin{align}
	\Delta \phi_\text{SU(1,1)}(\phi)	&= \sqrt{\tfrac{1}{8}\text{csch}^4(2r)\left\{ 4 \left[ 2 (\Delta n)^2 - \sin^2\left( \phi/2\right)\right] \csc^2 \phi + \cosh(8r) \sec^2(\phi/2) \right\} - 1}.
\end{align}
This has an optimal operating point of 
\begin{equation}
	\phi_\text{opt} = 2\sin^{-1} \left( \sqrt{ \frac{2 \Delta n}{2 \Delta n + \sqrt{2 \left[ 2(\Delta n)^2 + \cosh(8r) - 1 \right]}} }\right),
\end{equation}
which is in general nonzero, and therefore a minimum sensitivity of
\begin{align}
	\Delta \phi_\text{SU(1,1)}(\phi_\text{opt})	&= | \text{csch}(2r) | \sqrt{\left[ 1 + \frac{\Delta n}{2}\left(\text{csch}^2(2r)\Delta n  + \sqrt{4 \coth^2(2r) +  \text{csch}^4(2r)(\Delta n)^2} \right)\right]}.
\end{align}

\section{Hybrid atom-light pump-enhanced SU(1,1) interferometry}

The calculations below proceed similarly to those for the spinor BEC outlined above.

\subsection{Quantum Fisher Information}
We assume a general four-mode pure input state of the form
\begin{equation}
	|\psi_0 \rangle = \sum_{N_a, N_b =0}^\infty c_{N_a, N_b} |N_a,N_b,0,0 \rangle, \label{input_state_AL}
\end{equation}
where $|n_{a_0}, n_{b_0}, n_{a_1}, n_{b_1} \rangle$ is the four-mode Fock state with $n_{a_0}$ particles in atomic pump mode $\hat{a}_0$, $n_{b_0}$ particles in photonic pump mode $\hat{b}_0$, and $n_{a_1}$ and $n_{b_1}$ particles in atomic side mode $\hat{a}_1$ and photonic side mode $\hat{b}_1$, respectively. The process that transfers correlated pairs of particles from the two pumps to the side modes is described by the map
\begin{equation}
	|N_a,N_b,0, 0\rangle \mapsto \sum_{k \leq \min(N_a,N_b)} d_k(N_a, N_b) |N_a - k, N_b - k, k, k\rangle. \label{Fock_mappy_AL}
\end{equation}
The four-wave mixing Hamiltonian [Eq.~(3) from the main text] is a special case of this map. The quantum state Eq.~(\ref{input_state_AL}) is therefore mapped to
\begin{equation}
	|\psi_1 \rangle = \sum_{N_a, N_b =0}^\infty \sum_{n \leq \min(N_a,N_b)} C_n(N_a, N_b) |N_a-n,N_b-n, n,n\rangle,
\end{equation}
where $C_n(N) \equiv d_{N_a, N_b} c_n(N_a, N_b)$. This state is then pumped-up by separately interfering the atomic and photonic modes via the beamsplitting operations:
\begin{subequations}
	\begin{align}
		\hat{a}_0(\theta)	&= \hat{a}_0 \cos \theta - i e^{i \vartheta} \hat{a}_1 \sin \theta, \\
		\hat{a}_1(\theta)	&= \hat{a}_1 \cos \theta - i e^{-i \vartheta} \hat{a}_0 \sin \theta, \\
		\hat{b}_0(\theta)	&= \hat{b}_0 \cos \theta - i e^{i \vartheta} \hat{b}_1 \sin \theta, \\
		\hat{b}_1(\theta)	&= \hat{b}_1 \cos \theta - i e^{-i \vartheta} \hat{b}_0 \sin \theta.
	\end{align}
\end{subequations}
For simplicity, we have assumed that the beamsplitting angles and phases for the atoms and photons are identical. The side modes $\hat{a}_1$ and $\hat{b}_1$ then undergo a phase shift $\phi/2$ corresponding to the unitary $\hat{U}_{\phi} = \exp(-i \phi \hat{N}_s/2)$, where $\hat{N}_s = \hat{a}_1^\dag \hat{a}_1 + \hat{b}_1^\dag \hat{b}_1$ is the number sum operator. The QFI is given by Eq.~(\ref{QFI_expression}) with $\hat{N}_s(\theta) = \hat{a}_1^\dag(\theta) \hat{a}_1(\theta) + \hat{b}_1^\dag(\theta) \hat{b}_1(\theta)$. Explicitly,
\begin{subequations}
\begin{align}
	\langle \hat{N}_s(\theta) \rangle 	&= \langle \hat{N}_{0} \rangle \sin^2 \theta + \langle \hat{N}_s \rangle \cos^2 \theta, \\
	\langle \hat{N}_s(\theta)^2 \rangle 	&= \langle \hat{N}_0^2 \rangle \sin^4 \theta + \langle \hat{N}_s^2 \rangle \cos^4 \theta + \left( \langle \hat{N}_0 \rangle + \langle \hat{N}_s \rangle + 3 \langle \hat{N}_0 \hat{N}_s \rangle - 2 A(\vartheta) \right) \cos^2 \theta \sin^2 \theta,
\end{align}
\end{subequations}
where $\hat{N}_0 = \hat{a}_0^\dag \hat{b}_0 + \hat{b}_0^\dag \hat{a}_0$, $A(\vartheta) = \exp(-i \vartheta)\langle \hat{a}_0 \hat{b}_0 \hat{a}_1^\dag \hat{b}_1^\dag \rangle + h.c.$, and the above expectations are all taken with respect to $|\psi_1\rangle$. The QFI is therefore
\begin{align}
	\mathcal{F}(\theta)	&= \text{Var}(\hat{N}_0) \sin^4 \theta +  \text{Var}(\hat{N}_s) \cos^4 \theta \notag \\
						&+ \left[\langle \hat{N}_0 \rangle \left(\langle \hat{N}_s \rangle + 1 \right) + \langle \hat{N}_s \rangle +  3\text{Cov}(\hat{N}_0, \hat{N}_s)  - 2A(\vartheta)\right] \cos^2 \theta \sin^2 \theta.
	 \label{QFI_pure_AL}
\end{align}
The QFI depends on both the total number of pump particles (via the operator $\hat{N}_0$) and the \emph{fraction} of particles in mode $\hat{a}_0$ compared with $\hat{b}_0$ (via $A(\vartheta)$).
 
\subsubsection{Optimal beamplitting parameters}
Since
\begin{align}
	\frac{\partial \mathcal{F}}{\partial \vartheta}	&=  \frac{i}{2}\left( e^{-2i \vartheta}\langle \hat{a}_0 \hat{b}_0 \hat{a}_1^\dag \hat{b}_1^\dag \rangle - e^{2 i \vartheta}\langle \hat{a}_0^\dag \hat{b}_0^\dag \hat{a}_1 \hat{b}_1 \rangle \right) \sin^2(2\theta),
\end{align}
there exist two critical values for the beamsplitter phase $\vartheta$ (modulo $\pi$):
\begin{equation}
	\vartheta_\pm = \frac{i}{2} \ln \left( \pm \sqrt{\frac{\langle \hat{a}_0^\dag \hat{b}_0^\dag \hat{a}_1 \hat{b}_1 \rangle}{\langle \hat{a}_0 \hat{b}_0 \hat{a}_1^\dag \hat{b}_1^\dag \rangle}}\right).  
\end{equation}
The beamsplitter phase only affects the QFI via $A(\vartheta)$; here $A(\vartheta_\pm) = \pm 2 |\langle \hat{a}_0 \hat{b}_0 \hat{a}_1^\dag \hat{b}_1^\dag \rangle |$. Since $2| \langle \hat{a}_0 \hat{b}_0 \hat{a}_1^\dag \hat{b}_1^\dag \rangle | \geq 0$, inspection of the QFI Eq.~(\ref{QFI_pure_AL}) reveals that $\vartheta_-$ maximizes the QFI. 

We can similarly determine the optimal beamsplitter angle:
\begin{align}
	\frac{\partial \mathcal{F}}{\partial \theta} 	&= \Big[ \text{Var}(\hat{N}_0) + \text{Var}(\hat{N}_s) + \Big( \text{Var}(\hat{N}_s) - \text{Var}(\hat{N}_0) + 3 \text{Cov}(\hat{N}_0, \hat{N}_s) \notag \\
					&  + \langle \hat{N}_s \rangle + \langle \hat{N}_0\rangle(\langle \hat{N}_s \rangle + 1) -  2 A(\vartheta_-) \Big) \cos(2\theta) \Big] \sin(2\theta).
\end{align}
Therefore, the critical points are $\theta_\text{opt} = 0, \pi/2$ and $\theta_c^{(x_c)} = \cos^{-1}(x_c)/2$ for
\begin{equation}
	x_c = -\frac{\text{Var}(\hat{N}_0) + \text{Var}(\hat{N}_s)}{\text{Var}(\hat{N}_s) - \text{Var}(\hat{N}_0) + 3 \text{Cov}(\hat{N}_0, \hat{N}_s) + \langle \hat{N}_s \rangle + \langle \hat{N}_0\rangle(\langle \hat{N}_s \rangle + 1) -  2 A(\vartheta_-)}.
\end{equation}
This final critical point only exists if $|x_c| < 1$. Thus,
\begin{subequations}
\begin{align}
	\mathcal{F}(0) 		&= \text{Var}(\hat{N}_s), \\
	\mathcal{F}(\pi/2)		&= \text{Var}(\hat{N}_0), \\
	\mathcal{F}(\theta_c^{(x_c)})	&= \tfrac{1}{4} \Big[ \text{Var}(\hat{N}_0) (1 - x_c)^2 + \text{Var}(\hat{N}_s) (1 + x_c)^2 \notag \\
			&+ \left( 3 \langle \hat{N}_0 \hat{N}_s\rangle + \langle \hat{N}_s \rangle - \langle \hat{N}_0 \rangle\left( 2 \langle \hat{N}_0 \rangle - 1 \right) -  2 A(\vartheta_-)\right) \left(1 - x_c^2 \right) \Big],
\end{align}
\end{subequations}
and
\begin{equation}
	\mathcal{F}(\theta_\text{opt}) = \begin{cases}
							\max \left\{ \mathcal{F}(0), \mathcal{F}(\pi/2), \mathcal{F}(\theta_c^{(x_c)})	\right\}, & |x_c| < 1 \\
							\max \left\{ \mathcal{F}(0), \mathcal{F}(\pi/2) \right\}, & |x_c| \geq 1
							\end{cases}
\end{equation}

\subsubsection{Undepleted pump regime}

We assume that both pump modes remain in coherent states such that
\begin{subequations}
\begin{align}
	\langle \hat{N}_0 \rangle		&= \overline{N} - \mathcal{N}_s, \\
	\langle \hat{N}_0^2 \rangle	&= \langle \hat{N}_{a_0}^2 \rangle + 2 \langle \hat{N}_{a_0} \hat{N}_{b_0} \rangle + \langle \hat{N}_{b_0}^2 \rangle = (\overline{N}-\mathcal{N}_s+1)(\overline{N}-\mathcal{N}_s), \\
	\langle \hat{N}_s^2 \rangle	&= 2 \mathcal{N}_s (\mathcal{N}_s + 1), \\
	\langle \hat{N}_0 \hat{N}_s \rangle	&= (\overline{N} - \mathcal{N}_s) \mathcal{N}_s, \\
	 \langle \hat{a}_0 \hat{b}_0 \hat{a}_1^\dag \hat{b}_1^\dag \rangle	&= \frac{i}{2} e^{i(\vartheta_{p,a} + \vartheta_{p,b} + \vartheta_\text{sq})} \sqrt{\mathcal{N}_s(\mathcal{N}_s+2) \left(\langle \hat{N}_{a_0} \rangle - \tfrac{1}{2}\mathcal{N}_s\right) \left(\langle \hat{N}_{b_0} \rangle - \tfrac{1}{2}\mathcal{N}_s\right) },
\end{align}
\end{subequations}
where $\overline{N}$ is the average \emph{total} number of particles in the pump modes (satisfying $\overline{N} = \langle \hat{N}_0 \rangle + \langle \hat{N}_s \rangle$), $\vartheta_{p,a}$ ($\vartheta_{p,b}$) is the phase of the atomic (photonic) pump coherent state, $\hat{N}_{a_0} = \hat{a}_0^\dag \hat{a}_0$, and $\hat{N}_{b_0} = \hat{b}_0^\dag \hat{b}_0$. This is consistent with an outcoupling process {$\hat{U}_\text{PA} = \exp\big[-i r (e^{i \vartheta_\text{sq}} \hat{a}_1 \hat{b}_1 + h.c. )\big]$} with $\mathcal{N}_s \equiv \langle \hat{N}_s \rangle =  2 \sinh^2 r$. Equation~(\ref{QFI_pure_AL}) reduces to
\begin{multline}
	\mathcal{F}(\theta)	= \left( \overline{N} - \mathcal{N}_s \right) \sin^4 \theta + \mathcal{N}_s (\mathcal{N}_s + 2) \cos^4 \theta \\
						+ \Big( \overline{N} + \mathcal{N}_s \left( \overline{N} - \mathcal{N}_s \right) - \sqrt{\mathcal{N}_s(\mathcal{N}_s+2) (2\langle \hat{N}_{a_0} \rangle - \mathcal{N}_s) (2\langle \hat{N}_{b_0} \rangle - \mathcal{N}_s) } \sin \nu \Big)\cos^2\theta \sin^2 \theta,
\end{multline}
where $\nu \equiv 2 \vartheta - \vartheta_{p,a} - \vartheta_{p,b} - \nu_\text{sq}$. The maximum QFI occurs at $\nu = 3 \pi / 2$ and for beamsplitter angle $\theta_\text{opt} = 0, \pi/2$ or $\theta_c^{(x_c)} = \cos^{-1}(x_c)/2$, where
\begin{equation}
	x_c = \frac{\overline{N} - \mathcal{N}_s \left( \mathcal{N}_s + 3 \right)}{\mathcal{N}_s \left( 2 \mathcal{N}_s + 1 - \overline{N}\right) + \sqrt{\mathcal{N}_s(\mathcal{N}_s+2) (2\langle \hat{N}_{a_0} \rangle - \mathcal{N}_s ) (2\langle \hat{N}_{b_0} \rangle - \mathcal{N}_s ) } \sin \nu}.
\end{equation}
Assuming optimal $\nu$, $\theta_c^{(x_c)}$ exists and $\overline{N} \gg 1$, an asymptotic expansion of $\theta_c^{(x_c)}$ in powers of $1/\overline{N}$ yields
\begin{equation}
	\theta_c^{(x_c)} = \frac{1}{4} \left( \pi + 2 \text{csc}^{-1}\left( \mathcal{N}_s + \frac{2\sqrt{\mathcal{N}_s(\mathcal{N}_s+2)n_f}}{1 + n_f}\right) \right) + \mathcal{O}\left( 1 / \overline{N} \right), \label{approx_theta_c_AL}
\end{equation}
where $n_f \equiv \langle \hat{N}_{a_0} \rangle / \langle \hat{N}_{b_0} \rangle$. Since $\text{csc}^{-1}(x)$ is undefined for $|x| < 1$, Eq.~(\ref{approx_theta_c_AL}) implies that the critical point $\theta_c^{(x_c)}$ only exists for $\mathcal{N}_s > \left[ 1 + n_f (n_f + 6) - 2 \sqrt{n_f(3+n_f)(1 + 3 n_f)}\right] / (n_f-1)^2$. For sufficiently large $\mathcal{N}_s$ (e.g. $\mathcal{N}_s \gtrsim 10$ for $n_f = 1$), $\theta_c^{(x_c)} \approx \pi / 4$. The QFI is given by
\begin{subequations}
\begin{align}
	\mathcal{F}(\pi/2)	&= \overline{N} - \mathcal{N}_s, \\
	\mathcal{F}(\theta_c^{(x_c)})	&= \tfrac{1}{4}\Big[ \left( \overline{N} - \mathcal{N}_s\right) (x_c - 1)^2 + \mathcal{N}_s (\mathcal{N}_s + 2) (x_c + 1)^2 \notag \\
					& +\Big( \overline{N} + \mathcal{N}_s (\overline{N} - \mathcal{N}_s) + \sqrt{\mathcal{N}_s(\mathcal{N}_s+2) \left(2\langle \hat{N}_{a_0} \rangle - \mathcal{N}_s\right) \left(2\langle \hat{N}_{b_0} \rangle - \mathcal{N}_s\right) } \Big)(1-x_c^2) \Big] \notag \\
					&= \frac{\left[ (1+n_f)\cosh(2r) + 2\sqrt{n_f} \sinh(2r)\right]^2}{8(1+n_f)(1+n_f + 2\sqrt{n_f}\coth r)\sinh^2 r} \overline{N} + \mathcal{O}\left(\overline{N}^0\right).
\end{align}
\end{subequations}
This is reported in Eq.~(9) of the main text.

\subsection{Phase sensitivity for number-sum measurement in undepleted pump regime}
We assume both pumps are initially in coherent states - i.e. $|\psi_0 \rangle = |\alpha_0 e^{i \vartheta_{p,a}}, \beta_0 e^{i \vartheta_{p,b}}, 0, 0 \rangle$ where $|\alpha_0|^2$ and $|\beta_0|^2$ are the average number of pump atoms and pump photons, respectively. The evolution through the interferometer proceeds as:
\begin{enumerate}
	\item Parametric amplification, described by $\hat{U}_\text{PA}(r) = \exp[-i r ( e^{i \vartheta_\text{sq}} \hat{a}_1 \hat{b}_1 + h.c.)]$:
	\begin{subequations}
	\begin{align}
		\hat{a}_1^{(1)}(r)	&= \hat{a}_1 \cosh r - i \hat{b}_1^\dag \sinh r, \\
		\hat{b}_1^{(1)}(r)	&= \hat{b}_1 \cosh r - i \hat{a}_1^\dag \sinh r.
	\end{align} 
	\end{subequations}
	The coherent pump amplitudes evolve to $|\alpha(r)|^2 = |\alpha_0|^2 - \sinh^2 r$ and $|\beta(r)|^2 = |\beta_0|^2 - \sinh^2 r$.
	\item {The atoms and light undergo the following beamsplitting operations:
	\begin{subequations}
	\begin{align}
		\hat{a}_0^{(2)}(r, \theta)	&= \hat{a}_0 \cos \theta - i e^{i \vartheta} \hat{a}_1^{(1)}(r) \sin \theta, \\
		\hat{a}_1^{(2)}(r, \theta)	&= \hat{a}_1^{(1)}(r) \cos \theta - i e^{-i \vartheta} \hat{a}_0 \sin \theta, \\
		\hat{b}_0^{(2)}(r, \theta)	&= \hat{b}_0 \cos \theta - i e^{i \vartheta} \hat{b}_1^{(1)}(r) \sin \theta, \\
		\hat{b}_1^{(2)}(r, \theta)	&= \hat{b}_1^{(1)}(r) \cos \theta - i e^{-i \vartheta} \hat{b}_0 \sin \theta.
	\end{align}
\end{subequations}}
	\item The unitary $\hat{U}(\phi) = \exp(-i \phi \hat{N}_s / 2)$ shifts the phase of the two side modes by $\phi/2$ relative to the pump mode:
\begin{equation}
	\hat{a}_1^{(3)}(r, \theta, \phi) = \hat{a}_1^{(2)}(r, \theta)  e^{-i \phi / 2}; \qquad \hat{b}_1^{(3)}(r, \theta, \phi) = \hat{b}_1^{(2)}(r, \theta)  e^{-i \phi / 2}.
\end{equation}
	\item The second set of beamsplitters with angle $-\theta$:
	\begin{subequations}
	\begin{align}
		\hat{a}_0^{(4)}(r, \theta, \phi)	&= \hat{a}_0^{(2)}(r, \theta) \cos \theta + i e^{i \vartheta} \hat{a}_1^{(3)}(r, \theta, \phi) \sin \theta, \\
		\hat{a}_1^{(4)}(r, \theta, \phi)	&= \hat{a}_1^{(3)}(r, \theta, \phi) \cos \theta + i e^{-i \vartheta} \hat{a}_0^{(2)}(r, \theta) \sin \theta, \\
		\hat{b}_0^{(4)}(r, \theta, \phi)	&= \hat{b}_0^{(2)}(r, \theta) \cos \theta + i e^{i \vartheta} \hat{b}_1^{(3)}(r, \theta, \phi) \sin \theta, \\
		\hat{b}_1^{(4)}(r, \theta, \phi)	&= \hat{b}_1^{(3)}(r, \theta, \phi) \cos \theta + i e^{-i \vartheta} \hat{b}_0^{(2)}(r, \theta) \sin \theta.
	\end{align}
	\end{subequations}
	\item Finally, a second parametric amplification $\hat{U}_\text{PA}(-r)$:
	\begin{subequations}
	\begin{align}
		\hat{a}_1^{(5)}(\theta, r, \phi)	&= \hat{a}_1^{(4)}(r, \theta, \phi) \cosh r + i \big[\hat{b}_1^{(4)}(r, \theta, \phi)\big]^\dag \sinh r, \\
		\hat{b}_1^{(5)}(\theta, r, \phi)	&= \hat{b}_1^{(4)}(r, \theta, \phi) \cosh r + i \big[\hat{a}_1^{(4)}(r, \theta, \phi)\big]^\dag \sinh r,
	\end{align} 
	\end{subequations}
\end{enumerate}
We take $\hat{\mathcal{S}} = \big[\hat{a}_1^{(5)}(\theta, r, \phi)\big]^\dag \hat{a}_1^{(5)}(\theta, r, \phi) + \big[\hat{b}_1^{(5)}(\theta, r, \phi)\big]^\dag \hat{b}_1^{(5)}(\theta, r, \phi)$ as our measurement signal. By expressing $\hat{\mathcal{S}}$ in terms of $\hat{a}_0$, $\hat{a}_1$, $\hat{b}_0$, and $\hat{b}_1$, and taking expectations with respect to the initial state:
\begin{align}
	\langle \hat{\mathcal{S}} \rangle &= 2 \cos^4 \theta \sin^2(\phi/2) \sinh^2 (2r) + \sin^2 (2\theta) \sin^2(\phi/4) \Big[ \left( |\alpha(r)|^2 + |\beta(r)|^2 \right) \cosh(2r) \notag \\
		&  + 2 \sinh^2 r - 2 \alpha(r) \beta(r) \sin\left( \nu + \phi/2\right)\sinh(2r) \Big], \\
		\partial_{\phi}\langle \hat{\mathcal{S}} \rangle	&= \Big[ \left( |\alpha(r)|^2 + |\beta(r)|^2 \right) \sin^2 \theta \sin(\phi/2) \cosh(2r) + 2 \left( \sin^2 \theta \sin(\phi/2)+ 2 \cos^2 \theta \sin \phi \cosh^2 r\right) \sinh^2r \notag \\
				&  - 4 \alpha(r) \beta(r) \sin^2 \theta \sin\left( \nu + 3 \phi / 4\right) \sin(\phi/4) \sinh(2r)  \Big] \cos^2 \theta  \\
		\langle \hat{\mathcal{S}}^2 \rangle	&= \frac{\phi^2}{8} \cos^2 \theta \Big\{ 4 \cos^2 \theta \cosh(4r) - 2(1 + \cos^2 \theta) \notag \\
									&+ 2 \sin^2 \theta \left[ \left( |\alpha(r)|^2 + |\beta(r)|^2+1\right) \cosh(2r) - 2 \alpha(r) \beta(r) \sin \nu \sinh(2r) \right]\Big\} + \mathcal{O}(\phi^4).
\end{align}
We can determine \emph{a posteriori} that the optimal phase sensitivity occurs at $\phi = 0$, and therefore
\begin{align}
	\Delta \phi_N	&= \frac{ \sec^2 \theta \sqrt{ 2 \cosh(4r) - 1 - \sec^2 \theta + \left[ \left( |\alpha(r)|^2 + |\beta(r)|^2+1\right) \cosh(2r) + 2 \alpha(r) \beta(r) \sinh(2r)\right] \tan^2 \theta }}{\left|  \cosh(4r) - 1 + \left[ \left( |\alpha(r)|^2 + |\beta(r)|^2+1\right)\cosh(2r) + 2 \alpha(r) \beta(r) \sinh(2r) - 1 \right] \tan^2 \theta \right|}, \notag \\
					&= 2|\text{csc}(2\theta)|\left[ \frac{1}{\sqrt{\eta(r) \overline{N}}} + \frac{\left(\frac{1 + n_f}{\sqrt{n_f}}\cosh r \sin \nu - 2 \sinh r\right) \sinh^3 r}{\left( \eta(r) \overline{N}\right)^{3/2}}\right] + \mathcal{O}(1 / \overline{N}^{5/2}),
\end{align}
where $\eta(r) \equiv \cosh(2r) - 2 \left[ \sqrt{n_f} / (1 + n_f) \right] \sin \nu \sinh(2r)$, $\overline{N} = |\alpha_0|^2 + |\beta_0|^2$, and $n_f = |\alpha_0|^2 / |\beta_0|^2$. Therefore, to leading order in $\overline{N}$ the optimal parameters are $\nu = 3 \pi / 2$ and $\theta = \pi/4$. 

\subsubsection{Insensitivity to detection noise}

For an uncertainty $\Delta n$ in particle detection, the phase sensitivity for parameters $\nu = 3 \pi /2$ and $\theta = \pi/4$ is
\begin{align}
	\Delta \phi 	&= \sqrt{\frac{(1+n_f)\sec^2(\phi/4)\left[ (1+n_f) B(r,\phi) + 2\sqrt{n_f} C(r,\phi) \right]}{\left[ (1+n_f) \cosh(2r) + 2 \sqrt{n_f}\left( 2\cos(\phi/2)-1\right)\sinh^2(2r) \right]^2 \overline{N}}} + \mathcal{O}(1/\overline{N}^{3/2}).
\end{align}
where
\begin{subequations}
\begin{align}
	B(r,\phi)	&\equiv 4 \cos^4(\phi/4) \cosh(2r) + 4 \sin^2(\phi/4) \cosh(4r) + \sin^2(\phi/2) \cosh(6r), \\
	C(r,\phi)	&\equiv \cos (\phi/2) \left[ 1 + \left(4 \cosh(2r)\right) \cosh(2r) \right]\sinh(2r) \notag \\
			&- 8 \left( \cos \phi + \cos (3 \phi/2) \cosh^2 r \right) \cosh r \sinh^3 r - \sinh (4r).
\end{align}
\end{subequations}
To leading order the optimal operating point remains $\phi = 0$, and the sensitivity is \emph{independent} of the number resolution - i.e. $\Delta \phi \approx  2 \exp(-r) / \sqrt{\overline{N}}$ for $n_f = 1$. 

\section{Truncated Wigner stochastic phase-space simulations}

The truncated Wigner (TW) stochastic phase-space simulation method and its many applications have been described in detail elsewhere 
[33--36]. In brief, the master equation governing the quantum state's evolution is mapped to a partial differential equation (PDE) for the Wigner quasiprobability distribution 
[37, 38]. Once third- and higher-order derivatives are truncated (an uncontrolled approximation, but one that is typically valid provided the population per mode is not too great over the simulation time period 
[39], the PDE for the Wigner function takes the form of a Fokker-Planck equation, which can be efficiently simulated via a set of stochastic differential equations (SDEs). 

Specifically, the master equation for a spinor BEC undergoing spin-mixing and two-body losses is
\begin{equation}
	\partial_t \hat{\rho} = -\frac{i}{\hbar}[\hat{H}_\text{SMD}, \hat{\rho}] + \sum_{i,j= 0, \pm} \gamma_{i,j} \mathcal{D}[\hat{a}_i \hat{a}_j] \hat{\rho},
\end{equation}
where $\mathcal{D}[\hat{L}]\hat{\rho} \equiv \hat{L} \hat{\rho} \hat{L}^{\dag} - \tfrac{1}{2} \{ \hat{L}^{\dagger} \hat{L}, \hat{\rho} \}$ and $\gamma_{i,j}$ are two-body loss rates due to recombination between modes $\hat{a}_i$ and $\hat{a}_j$. This can be unravelled into the following set of Ito SDEs [40] 
\begin{subequations}
\label{TW_SMD}
\begin{align}
d\alpha_0	& = \left[-\frac{i}{\hbar}\left( 2 \kappa \alpha_+\alpha_-\alpha^*_0 \right) - \left(\gamma_{0,0}|\alpha_0|^2 + \frac{\gamma_{0,+}}{2}|\alpha_+|^2 + \frac{\gamma_{0,-}}{2}|\alpha_-|^2\right)\alpha_0 \right] dt \notag \\
 &+ \sqrt{\frac{\gamma_{0,+}}{2}}\alpha^*_+d \xi_{1}(t)  + \sqrt{\frac{\gamma_{0,-}}{2}}\alpha^*_- d\xi_{2}(t) + \sqrt{2\gamma_{00}}\alpha^*_0 d\xi_{3}(t) , \\
 d\alpha_\pm & = \left[-\frac{i}{\hbar}\left( \kappa \alpha^2_0\alpha^*_\mp \right) - \frac{\gamma_{0,\pm}}{2}|\alpha_0|^2\alpha_\pm \right] dt + \sqrt{\frac{\gamma_{0,\pm}}{2}}\alpha^*_0d\xi_{\pm}(t),
\end{align}
\end{subequations}
where the $d\xi_i$ are complex Wiener noises satisfying $\overline{d\xi_i^*(t)} = 0$ and $\overline{ d\xi_i^*(t) d \xi_j(t)} = \delta_{i,j} dt$. The complex amplitudes $\alpha_i$ correspond to the modes $\hat{a}_i$; formally, averages over the resulting stochastic variables correspond to symmetrically-ordered expectations, for example $ \overline{|\alpha_i|^2} = \langle \hat{a}_i^\dag \hat{a}_i + \hat{a}_i \hat{a}_i^\dag \rangle / 2$. The initial conditions for Eqs.~(\ref{TW_SMD}) are obtained by randomly sampling the Wigner distribution for a coherent state (for mode $\hat{a}_0$) or a vacuum state (for modes $\hat{a}_\pm$); explicitly $\alpha_0(0) = \sqrt{\overline{N}} + \eta_0$ and $\alpha_\pm(0) = \eta_\pm$ for independent Gaussian noises $\eta_i$ satisfying $\overline{\eta_i} = 0$ and $\overline{\eta_i \eta_j} = \delta_{i,j} / 2$ [41]. 

In Fig.~2(a) of the main text, pump depletion during the spin-mixing step was accounted for by numerically simulating Eqs.~(\ref{TW_SMD}) with the loss rates set to zero. This gave the required expectations needed to compute, for example, the QFI Eq.~(\ref{QFI_pure}). For the upper left panel of Fig.~3, for simplicity we assumed $\gamma_{i,j} = \gamma$ for all $i,j$. 

For our hybrid atom-light system, one-body losses from the pump modes during the Raman process were modelled using
\begin{equation}
	\partial_t \hat{\rho} = -\frac{i}{\hbar}[\hat{H}_\text{FWM}, \hat{\rho}] + (\gamma_{a_0} \mathcal{D}[\hat{a}_0] + \gamma_{b_0} \mathcal{D}[\hat{b}_0])\hat{\rho},
\end{equation}
which after the TW approximation was unravelled into the following set of SDEs:
\begin{subequations}
\begin{align}
 	d\alpha_0 & = \left(-\frac{i}{\hbar} \kappa \beta^*_0\beta_1\alpha_1 - \frac{\gamma_{a_0}}{2}\alpha_0\right) dt + \sqrt{\frac{\gamma_{a_0}}{2}}d\xi_a(t) , \\
 	d\alpha_1 & = -\frac{i}{\hbar} \kappa \beta^*_0\beta_1\alpha_0 dt, \\
 	d\beta_0 & = \left(-\frac{i}{\hbar} \kappa \alpha^*_0\alpha_1\beta_1 - \frac{\gamma_{b_0}}{2}\beta_0 \right) dt + \sqrt{\frac{\gamma_{b_0}}{2}}d\xi_b(t) , \\
 	d\beta_1 & =   -\frac{i}{\hbar} \kappa \alpha^*_0\alpha_1\beta_0 dt ,
\end{align}
\end{subequations}
where $\gamma_{a_0}$ and $\gamma_{b_0}$ are the single-body loss rates for the atomic and photonic modes, respectively, the operator correspondences are $\hat{a}_i \to \alpha_i$ and $\hat{b}_i \to \beta_i$, and $d\xi_i(t)$ are complex Wiener noises satisfying $\overline{d\xi_i^*(t)} = 0$ and $\overline{ d\xi_i^*(t) d \xi_j(t)} = \delta_{i,j} dt$. Again, for simplicity we set $\gamma_{a_0} = \gamma_{b_0}$.

\section{Absolute sensitivities for pumped-up SU(1,1) interferometry under losses }
Fig.~3 of the main text displays the \emph{relative} sensitivities for pumped-up SU(1,1) interferometry compared with conventional SU(1,1) interferometry. However, when one is interested in the regimes where pumped-up SU(1,1) interferometry surpasses the shot-noise limit, the \emph{absolute} sensitivity is the relevant metric. Although much of this data can be extracted directly from Fig.~3 of the main text (by virtue of the many well-established results from conventional SU(1,1) sensitivities), for convenience we have plotted these sensitivities below in Fig.~\ref{fig_losses_sup}.	

\begin{figure}%
\centering
\includegraphics[width=\columnwidth]{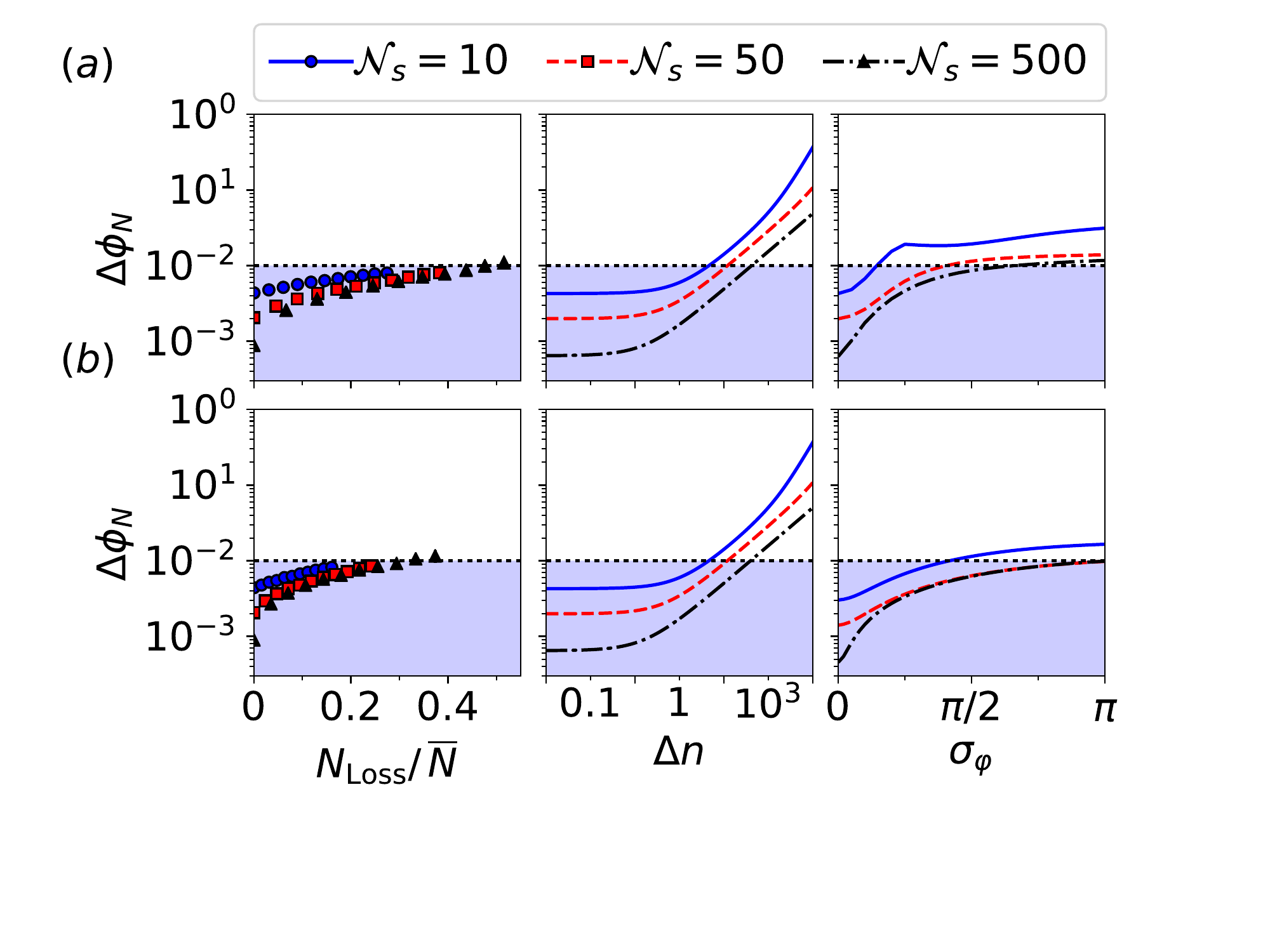}
\caption{Absolute sensitivities of pumped-up SU(1,1) interferometry in (a) a spinor BEC setup [top panels] and (b) a hybrid atom-light system [bottom panels]. Total particle number is $\overline{N} = 10^4$ and $n_f = 1$ for the hybrid atom-light system. All values plotted are for optimal $\phi$, optimal angle $\theta_\text{opt}$ and optimal phase $\vartheta_\text{opt}$. These show the dependence on: (Left) fraction of particles lost due to (a) two-body and (b) one-body losses, obtained via truncated Wigner simulations; (Middle) imperfect particle detection with number resolution $\Delta n$, obtained from semi-analytic calculations (see above); (Right) Gaussian phase-difference noise of variance $\sigma_\varphi^2$, obtained from analytic calculations with $\phi$ optimized numerically. The shaded region indicates the parameter regimes where pumped-up SU(1,1) surpasses the shot-noise limit. The side mode populations $\mathcal{N}_s = 10, 50$, and $500$ correspond to approximately $13.4$~dB, $20$~dB, and $30$~dB of squeezing, respectively. 
}
\label{fig_losses_sup}
\end{figure}

\end{widetext}

\end{document}